\documentclass[a4paper,5p]{elsarticle}

\journal{Computer Physics Communications}

\usepackage{graphicx}
\usepackage{bm}

\usepackage{color}
\usepackage{psfrag} 

\graphicspath{{figure_numerate/}{nuove_fig2013/}{fig_ene/}}

\newcommand{\ub}{{\bf u}}

\newcommand{\omegab}{\mbox{\boldmath $\omega$}}

\newcommand{\pdd}[2]{\frac{\partial #1}{\partial #2}} 

\begin{document}

\begin{frontmatter}

\title{Large-eddy simulation of hypersonic flows. Selective procedure to activate the sub-grid model only where small scale turbulence is present.} 

\author[a]{D.Tordella\corref{cor1}}
\ead{daniela.tordella@polito.it}
\author[a]{M.Iovieno}
\author[b]{S.Massaglia}
\author[b]{A.Mignone}
\address[a]{Politecnico di Torino, Dipartimento di Ingegneria Meccanica e Aerospaziale, Corso Duca degli Abruzzi 24, 10129 Torino, Italy}
\address[b]{Universit\`a di Torino, Dipartimento di Fisica Generale, via P.Giuria 1, 10125 Torino, Italy}

\cortext[cor1]{Corresponding author}
\date{-}

\begin{abstract}
 A new method for the localization of the regions where small scale  turbulent fluctuations are present in hypersonic flows is applied to the large-eddy simulation (LES) of a compressible turbulent jet with an initial Mach number equal to 5. The localization method used is called selective LES and is based on the exploitation of a scalar probe function $f$ which represents the magnitude of the \textit{stretching-tilting} term of the vorticity equation normalized with the enstrophy \cite{tim07}. For a fully developed turbulent  field of fluctuations, statistical analysis shows that the probability that $f$ is larger than 2 is almost zero, and, for any given threshold, it is larger if the flow is under-resolved. By computing the spatial field of $f$ in each  instantaneous realization  of the simulation it is possible to locate the regions where the magnitude of the normalized vortical stretching-tilting is anomalously high.
The sub-grid model is then introduced into the governing equations in such regions only.
The results of the selective LES simulation are compared with those of a standard LES, where the sub-grid terms are used in the whole domain, and with those of a standard Euler simulation with the same resolution. The comparison is carried out by assuming as reference field a higher resolution Euler simulation of the same jet. It is shown that the \textit{selective} LES modifies the dynamic properties of the flow to a lesser extent with respect to the classical LES. In particular, the prediction of the enstrophy, mean velocity and density distributions and of the energy and density spectra are substantially improved.
\end{abstract}


\end{frontmatter}
\section{Introduction : the small scale detection criterion}
Turbulent flows in many different physical and engineering applications have a Reynolds number so high that a direct numerical simulation of the Navier-Stokes equations (DNS) is not feasible.
The large-eddy simulation (LES) is a method in which the large scales of turbulence only are directly solved while the effects of the small-scale motions are modelled. 
The mass, momentum and energy equations are filtered in space in order to obtain the governing equation for the large scale motions.
The momentum and energy transport at the large-scale level due to the unresolved scales is represented by the so-called subgrid terms. Standard models for such terms, as, for example, the widely used Smagorinsky model, are based on the assumption that the unresolved scales are present in the whole domain and that turbulence is in equilibrium at subgrid scales (see, e.g., \cite{lilly,cc97}).
This hypothesis can be questionable in free, transitional and highly compressible turbulent flows where subgrid scales, that is fluctuations on a scale smaller than the space filter size, are not simultaneously present in the whole domain.
In such situations, subgrid models such as Smagorinsky's  overestimate the energy flow toward subgrid scales and, from the point of view of the large, resolved, scales, they appear as over-dissipative by exceedingly damping the large-scale motion.

For instance, simulation of astrophysical jets could suffer from such limitation. In this regard, any improvement of the LES methodology is opportune. Astrophysical flows occur in very large sets of spatial scales and velocities, are highly compressible (Mach number up to $10^2$) and have a Reynolds number which can exceed $10^{13}$, so that only the largest scales of the flow can be resolved even by the largest simulation in the foreseeable future. As a consequence, today, in this field, LES appears as a feasible simulation methods able to predict the unsteady system behaviour.

We have recently proposed a simple method to localize the regions where the flow is underresolved \cite{tim07}.
The criterion is based on the introduction of a local functional of vorticity and velocity gradients. The  regions where the fluctuations are unresolved are located by means of the scalar probe function \cite{tim07} which is based on the  vortical stretching-tilting   sensor:
\begin{equation}
f(\ub,\omegab) = \frac{\mid (\omegab - \overline{\omegab}) \cdot \nabla (\ub-
\overline{\ub}) \mid}{\mid \omegab-\overline{\omegab} \mid^2}
\label{f-DNSfluctuation}
\end{equation}
\noindent
where $\ub$ is the velocity vector, $\omegab = \nabla\times \ub$ is the vorticity vector and the overbar indicates the statistical average.
Function (\ref{f-DNSfluctuation}) is a normalized scalar form of the vortex-stretching term that represents the inertial generation of three dimensional vortical small scales inside the vorticity equation. When the flow is three dimensional and rich in small scales $f$ is necessarily different from zero, while, on the other hand, it is instead equal to zero in a two-dimensional vortical flow where the vortical stretching is absent. The mean flow is subtracted from the velocity and vorticity fields in order to consider the fluctuating part  of the field only.

\begin{figure}
 \centering
 \includegraphics[width=0.85\columnwidth]{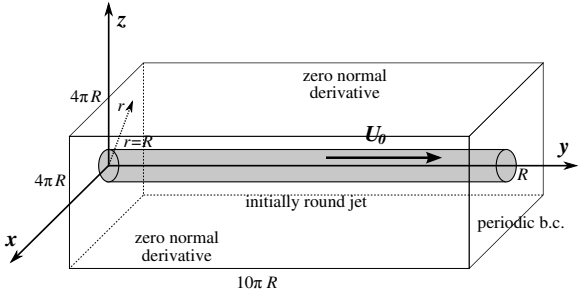}
 \caption{Scheme of the computational domain and boundary conditions. The initial condition is represented by the grey cylinder of radius $R$. The initial velocity field is a laminar parallel flow, with an initial Mach number equal to 5, perturbed by eight waves which have an amplitude equal to 5\%\ of the axis jet velocity and a wavelength from $1.25$ to $10$ times $\pi R$.}
 \label{fig.schema}
\end{figure}

\begin{figure}
\centering
\includegraphics[width=0.8\columnwidth,angle=0]{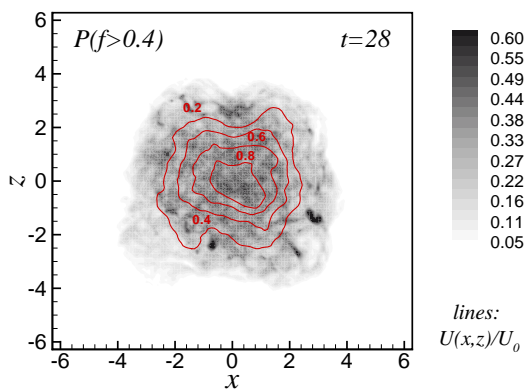}
\caption{Contour plots at $t=28$ of the probability that $f\ge t_\omega$ (see equation \ref{f-DNSfluctuation}) and thus the probability that subgrid terms are introduced in the selective LES balance equation by the localization procedure. The lines represent the points where the longitudinal velocity $\overline u/U_0$ is constant, where $U_0$ is the jet axis mean velocity. All data in this figure have been computed averaging on lines parallel to the jet axis. A three-dimensional animation which shows the time evolution of the underresolved regions where the subgrid terms are introduced in the selective LES can be seen in the supplemental material.}
\label{getto-prob}
\end{figure}

A priori test of the spatial distribution of functional test have been performed by computing the statistical distribution of $f$ in a fully resolved turbulent fluctuation field (DNS of a homogeneous and isotropic turbulent flow ($1024^3$, $Re_\lambda=230$, data from \cite{toschi})) and in some unresolved instances obtained by filtering this DNS field  on coarser grids (from $512^3$ to $64^3$).
It has been shown \cite{tim07}  that the probability that $f$ assumes values larger than a given threshold $t_\omega$ is always higher in the filtered fields and increases when the resolution is reduced. The difference between the probabilities in fully resolved and in filtered turbulence is maximum when $t_\omega$ is in the range $[0.4,0.5]$ for all resolutions. In such a range the probability $p(f\ge t_\omega)$ that $f$ is larger than $t_\omega$ in the less resolved field is about twice the probability in the DNS field. Furthermore, beyond this range this probability normalized over that of resolved DNS fields it is gradually increasing becoming infinitely larger. From that it is possible to introduce a threshold $t_\omega$ on the values of $f$, such that, when $f$ assumes larger values the field could be considered locally unresolved and should benefit from the local activation of the Large Eddy Simulation method (LES) by inserting a subgrid scale term in the motion equation. The values of this threshold is arbitrary, as there is no sharp cut, but it can be reasonably chosen as the one which gives the maximum difference between the probability $p(f\ge t_\omega)$ in the resolved and unresolved fields. This leads to $t_\omega\approx0.4$. Furthermore, it should be noted that the Morkovin hypothesis, stating that the compressibility effects do not have much influence on the turbulence dynamics, apart from varying the local fluid properties \cite{morkovin}, allows to apply the same value of the threshold in compressible and incompressible flows.

Such value of the threshold has been used to investigate the presence of regions with anomalously high values of the functional $f$, by performing a set of a priori tests on existing Euler simulations of the temporal evolution of a perturbed cylindrical hypersonic light jet with an initial Mach number equal to 5 and ten times lighter than the surrounding external ambient \cite{tim07}. When the effect of the introduction of subgrid scale terms in the transport equation is extrapolated from those a priori tests, they positively compare with experimental results and  show the convenience of the use of such a procedure \cite{tim07,br,ps02}.

In this paper we present large-eddy simulations of this temporal evolving jet, where the subgrid terms are selectively introduced in the transport equations by means of the local stretching criterion \cite{tim07}.
The aim is not to model a specific jet, but instead  to understand, from a physical point of view, the differences introduced by the presence of sub-grid terms in the under-resolved simulations of hypersonic jets.

Our localization procedure selects the regions where subgrid terms are applied and, as such, its effect could be considered equivalent to a model coefficient modulation, as the one obtained by the dynamic procedure \cite{germano} or by the use of improved eddy viscosity Smagorinsky-like models like Vreman's model \cite{vreman}, which gives a low eddy viscosity in non turbulent regions of the flow. However, it operates differently because it is completely uncoupled from the subgrid scale model used as, unlike the common practical implementations of the dynamic procedure, does not require ensemble averaging to prevent unstable eddy viscosity. Other alternatives, such as the approximate deconvolution model \cite{sak01}, are more complicated than the present selective procedure because involve filter inversion and the use of a dynamic relaxation term.
The computational overload of the selective filtering is modest and can make LES an affordable alternative to a higher resolution inviscid simulation: the selective LES increases the computing time of about one-third with respect to an Euler simulation, while the doubling of the resolution can increase the computing time by a factor of sixteen.


\section{Flow configuration and numerical method}

\begin{figure*}
\centering
\begin{tabular}{c}
\includegraphics[width=0.46\textwidth,angle=-90]{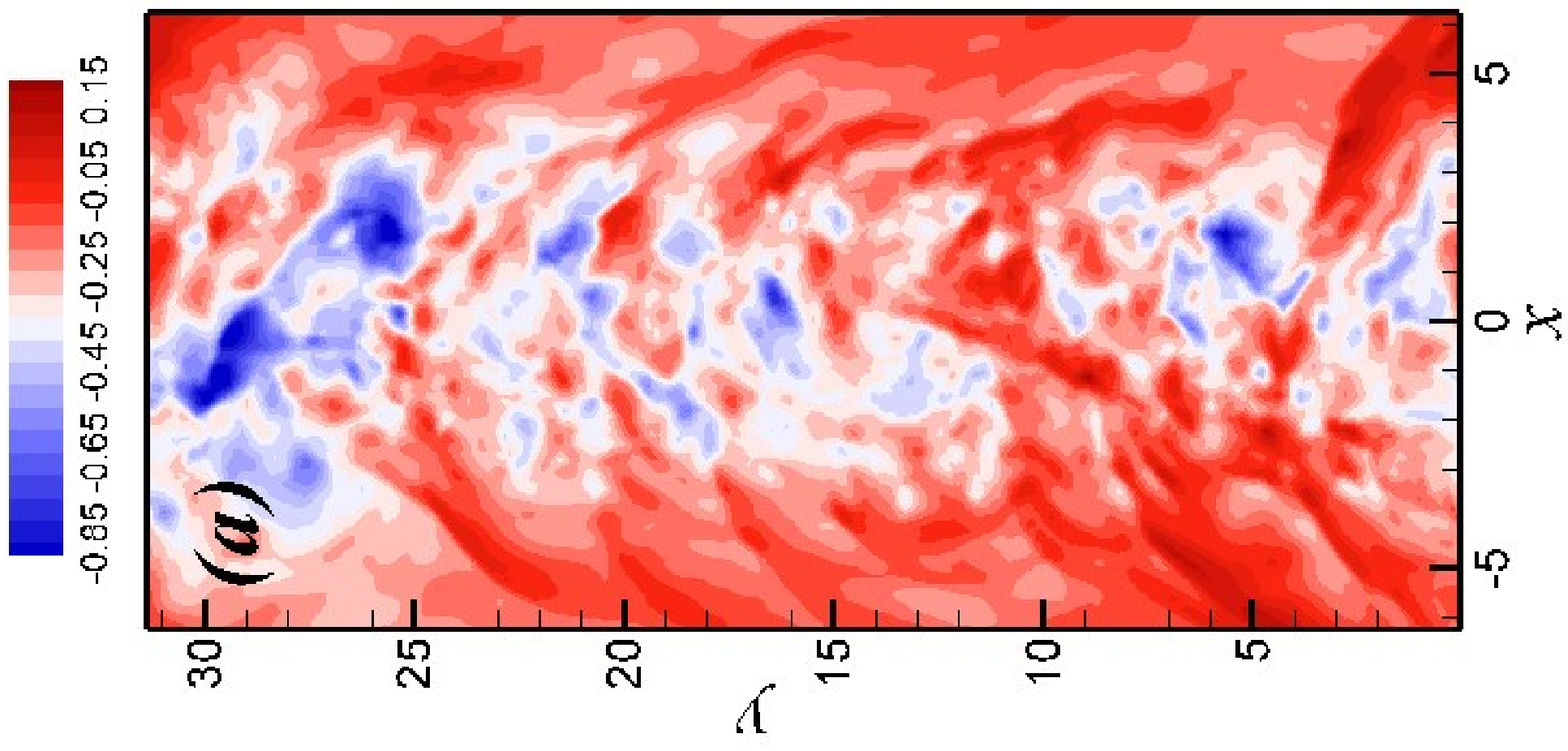} 
\includegraphics[width=0.46\textwidth,angle=-90]{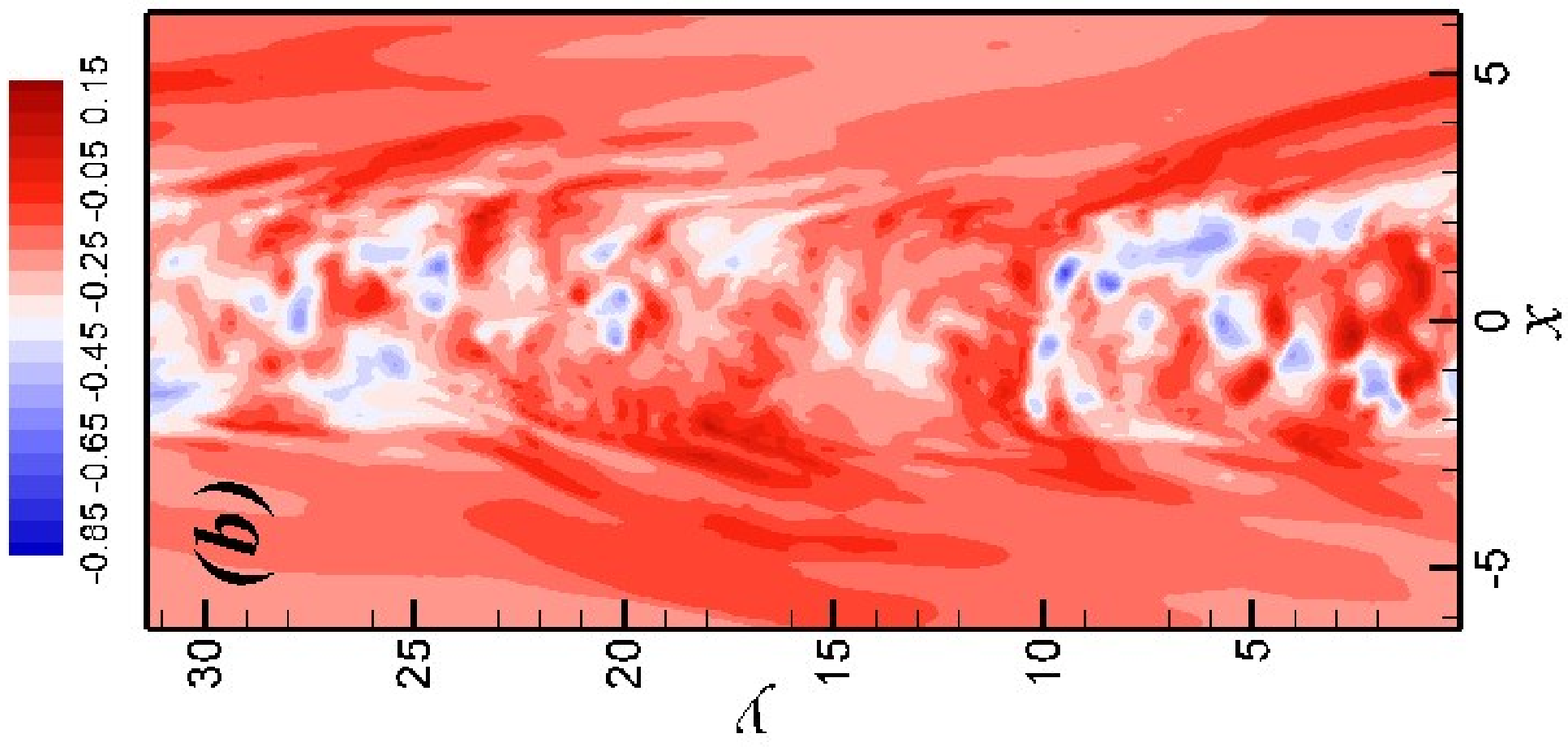} 
\includegraphics[width=0.46\textwidth,angle=-90]{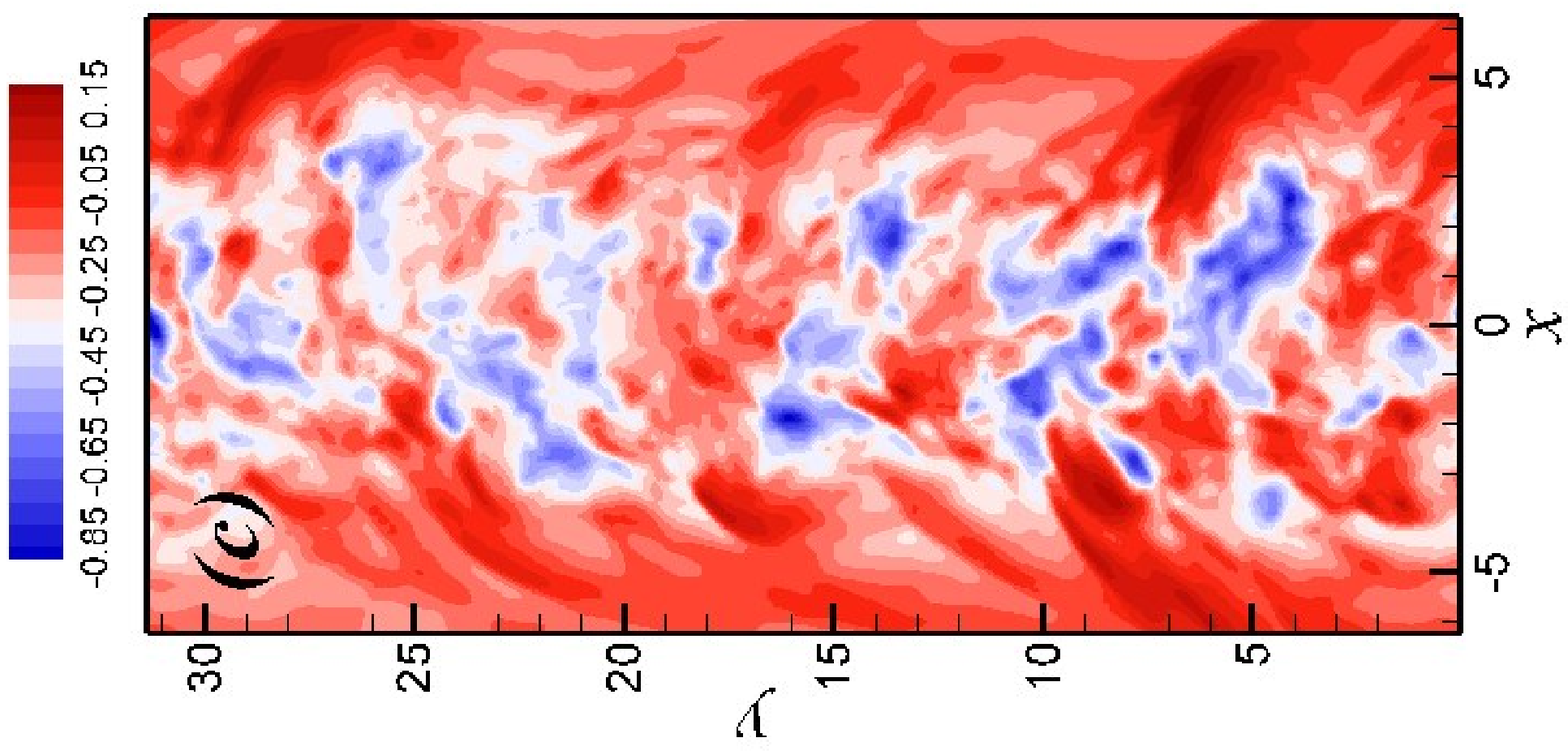} 
\includegraphics[width=0.46\textwidth,angle=-90]{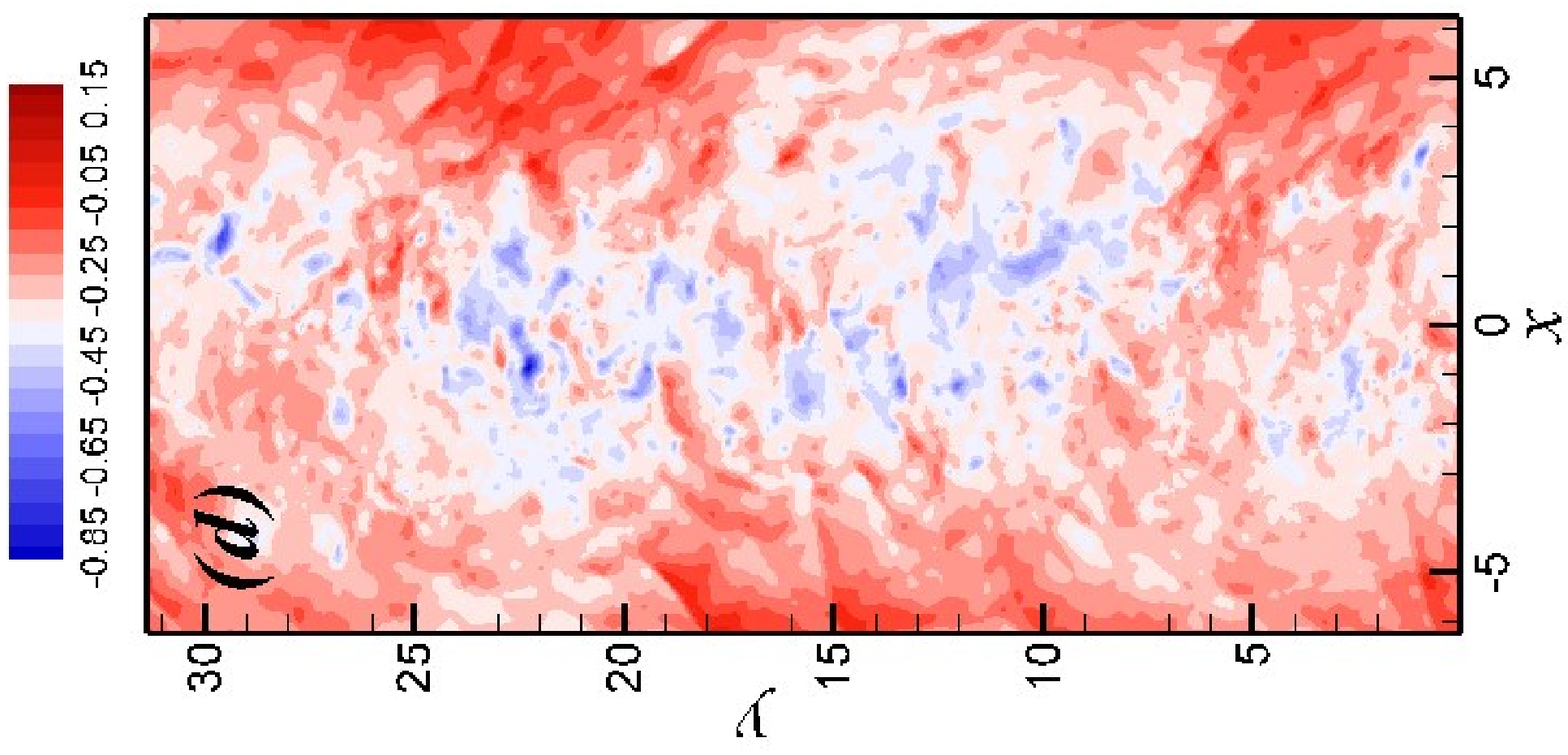}\\ 
\includegraphics[width=0.46\textwidth,angle=-90]{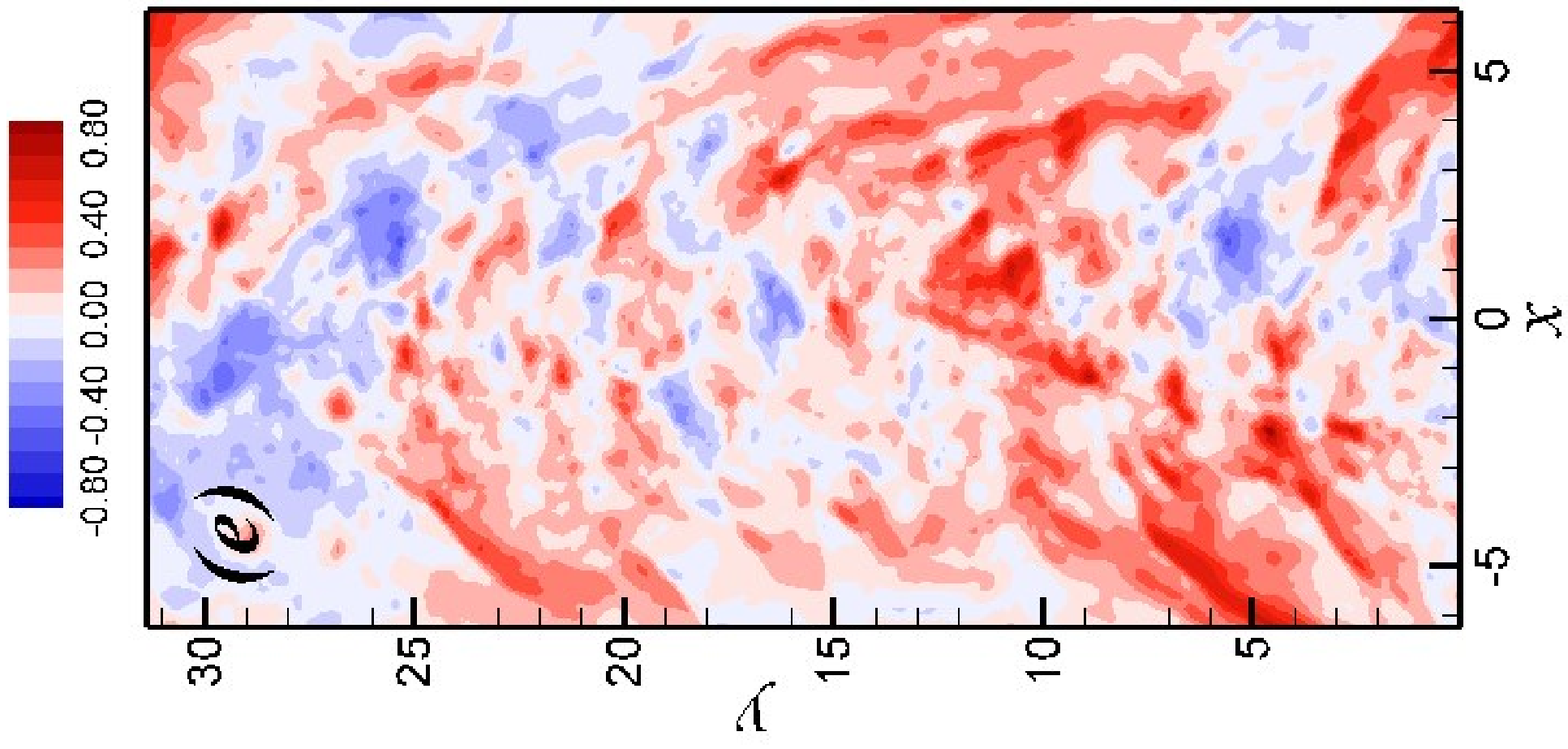} 
\includegraphics[width=0.46\textwidth,angle=-90]{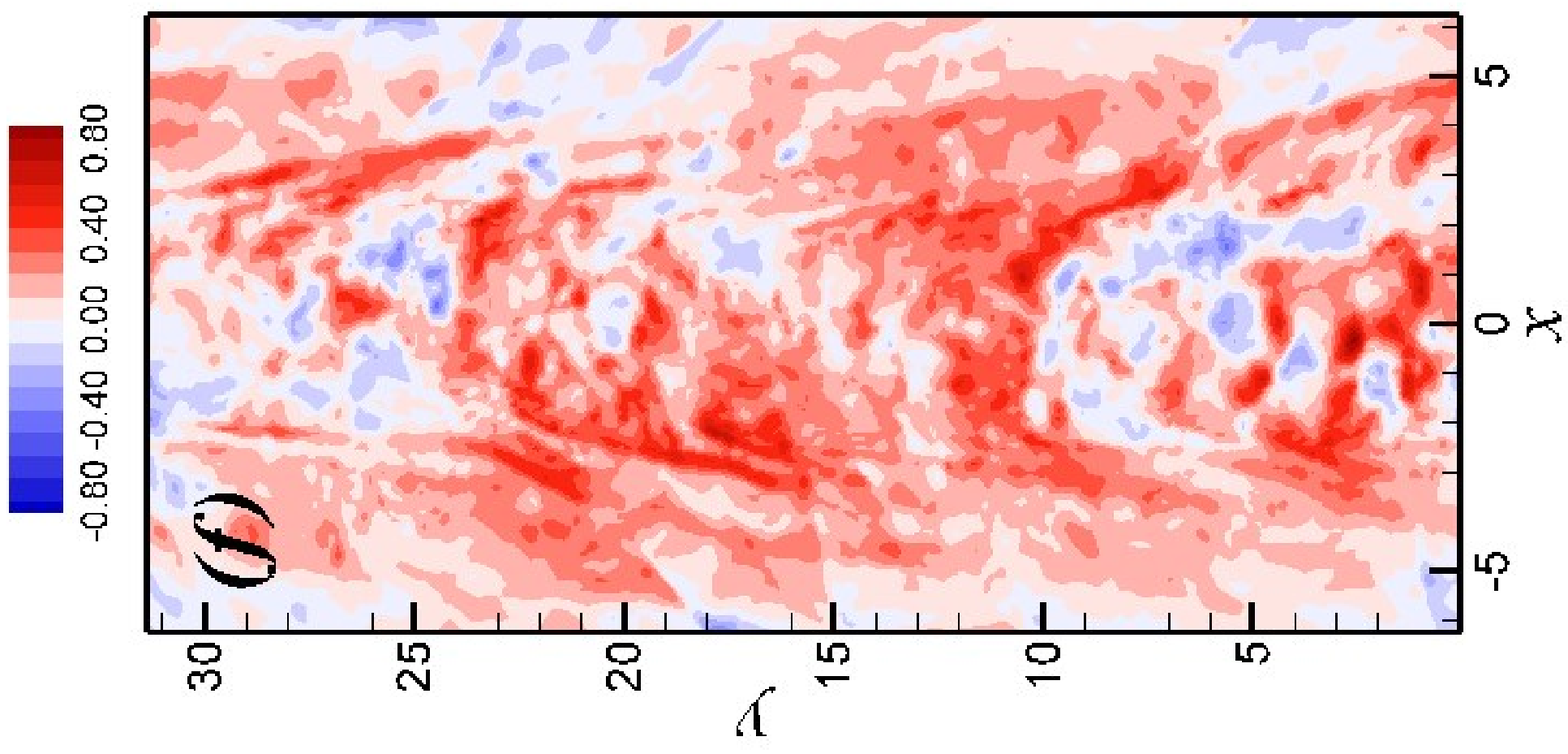} 
\includegraphics[width=0.46\textwidth,angle=-90]{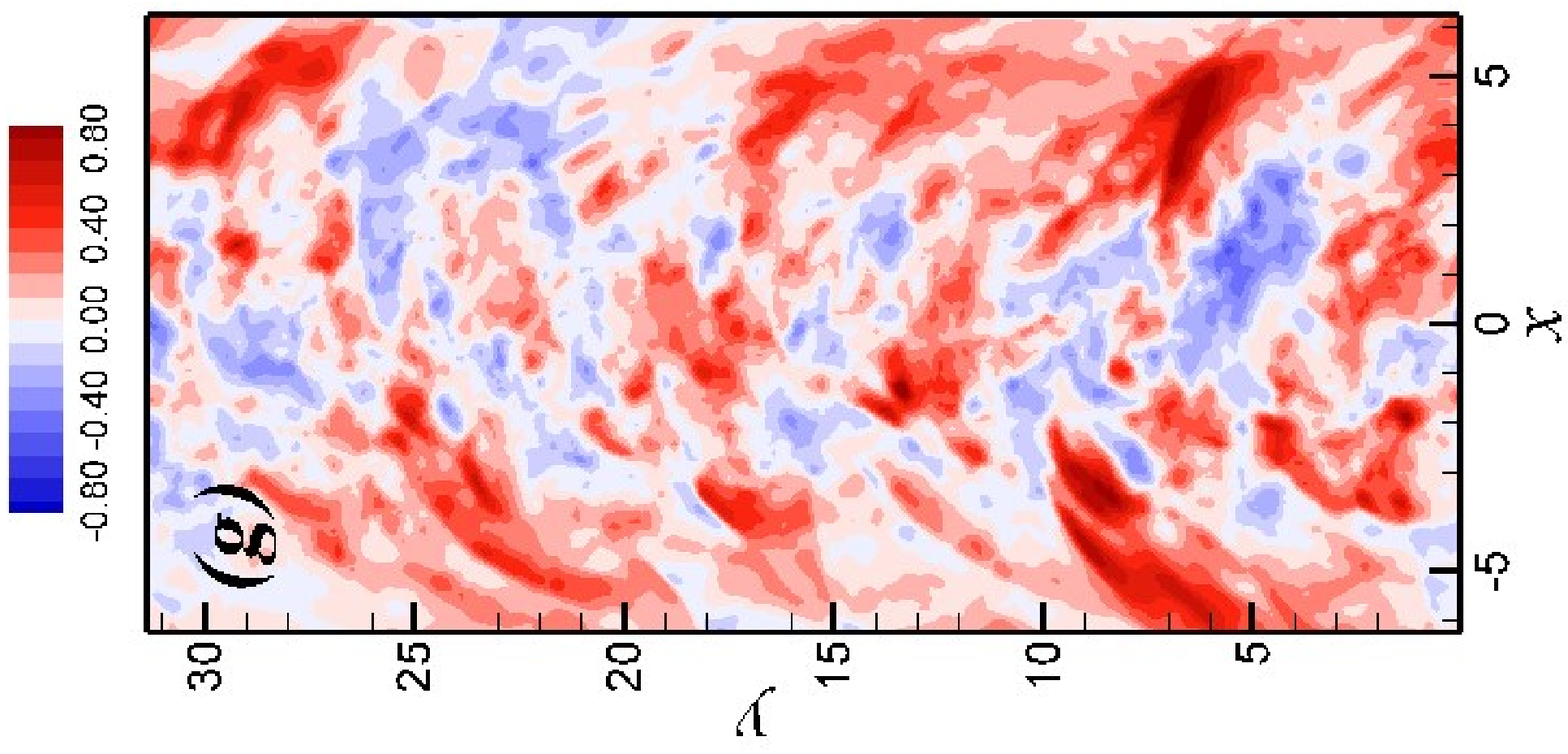} 
\end{tabular}
\caption{Top panel: pressure distribution in a longitudinal section at $t=32$: (a) selective LES, (b) standard LES, (c) low resolution pseudo-DNS, (d) higher resolution pseudo-DNS. The figures show the contour levels of $\log_{10} (p/p_0)$, the mean flow is from bottom to top.
Bottom panel: local difference between the lower resolution simulations and the higher resolution pseudo-DNS at $t/\tau=32$: (e) selective LES, (f) standard LES, (g) low resolution pseudo-DNS. The figures show the contour levels of $(p_{LES}-p_{DNS})/p_0$.}
\label{vis.pr}
\end{figure*}

\begin{figure*}
\centering
\begin{tabular}{c}
\includegraphics[width=0.46\textwidth,angle=-90]{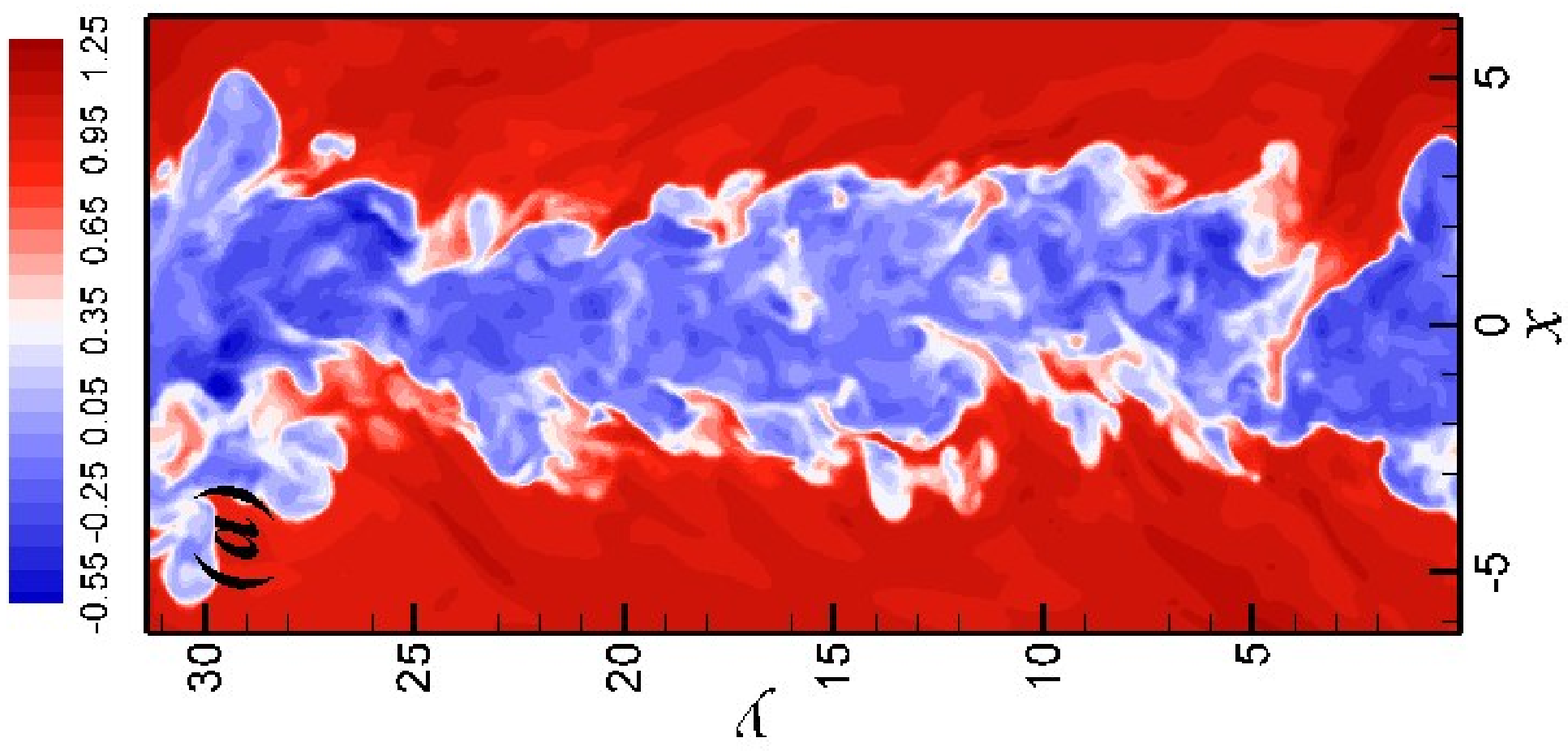} 
\includegraphics[width=0.46\textwidth,angle=-90]{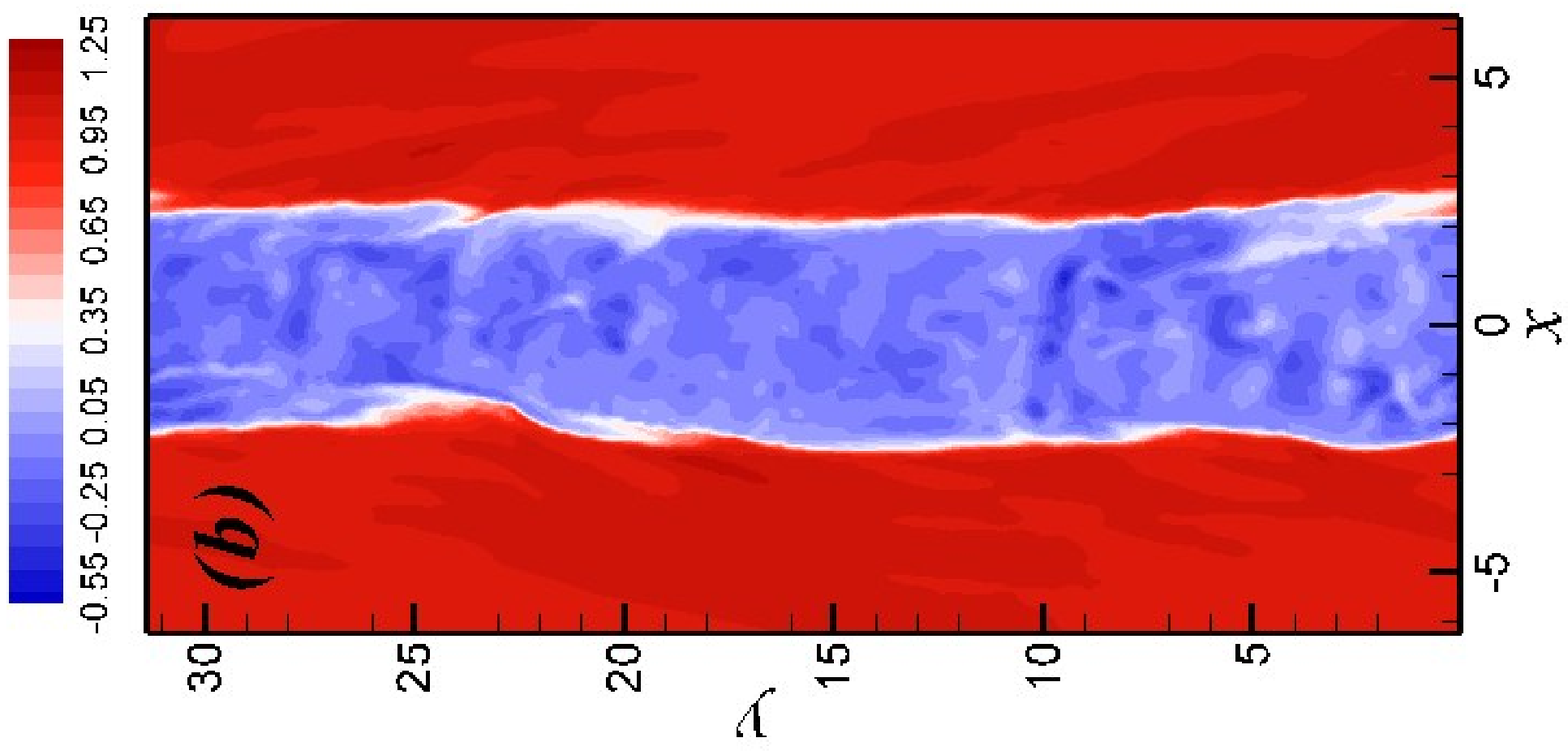} 
\includegraphics[width=0.46\textwidth,angle=-90]{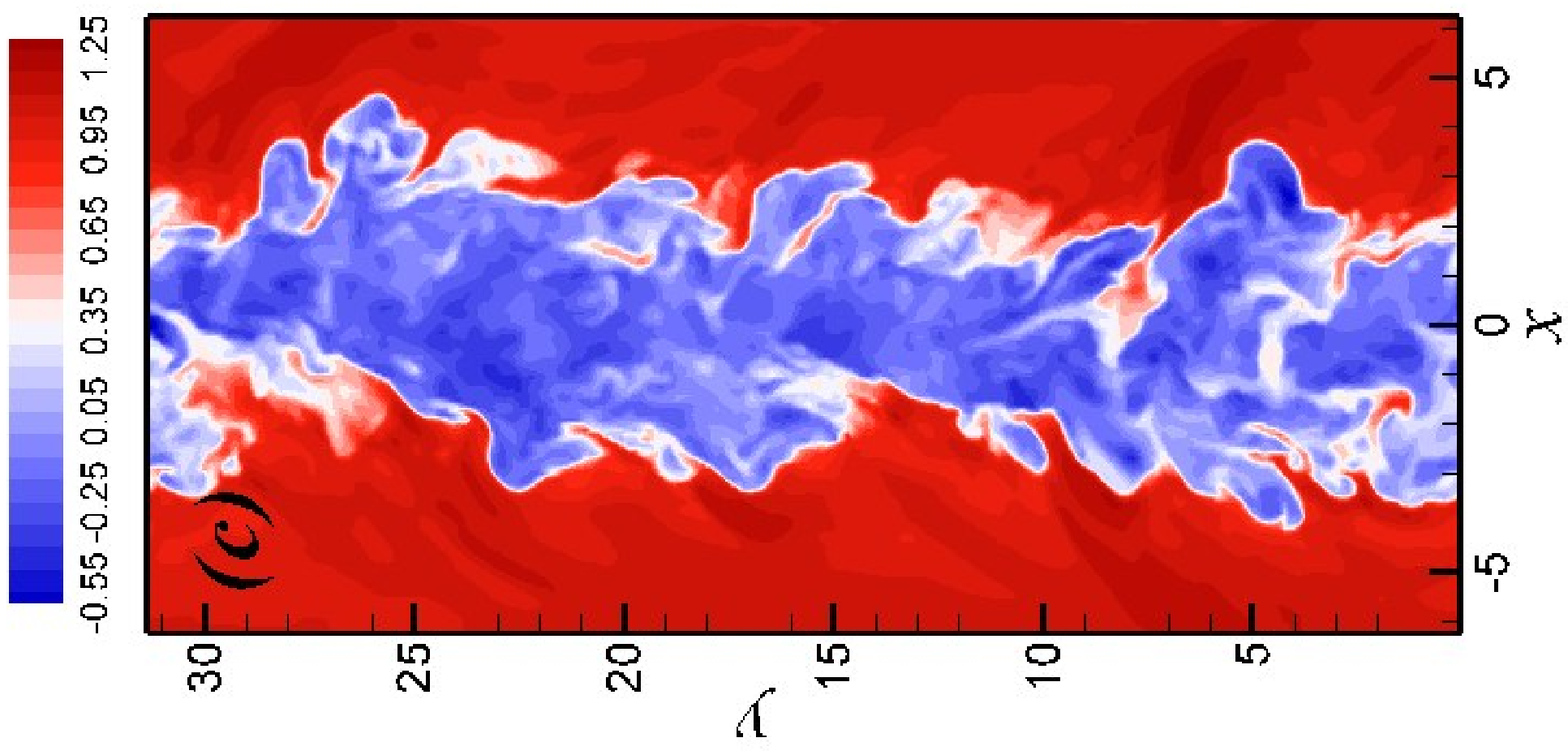} 
\includegraphics[width=0.46\textwidth,angle=-90]{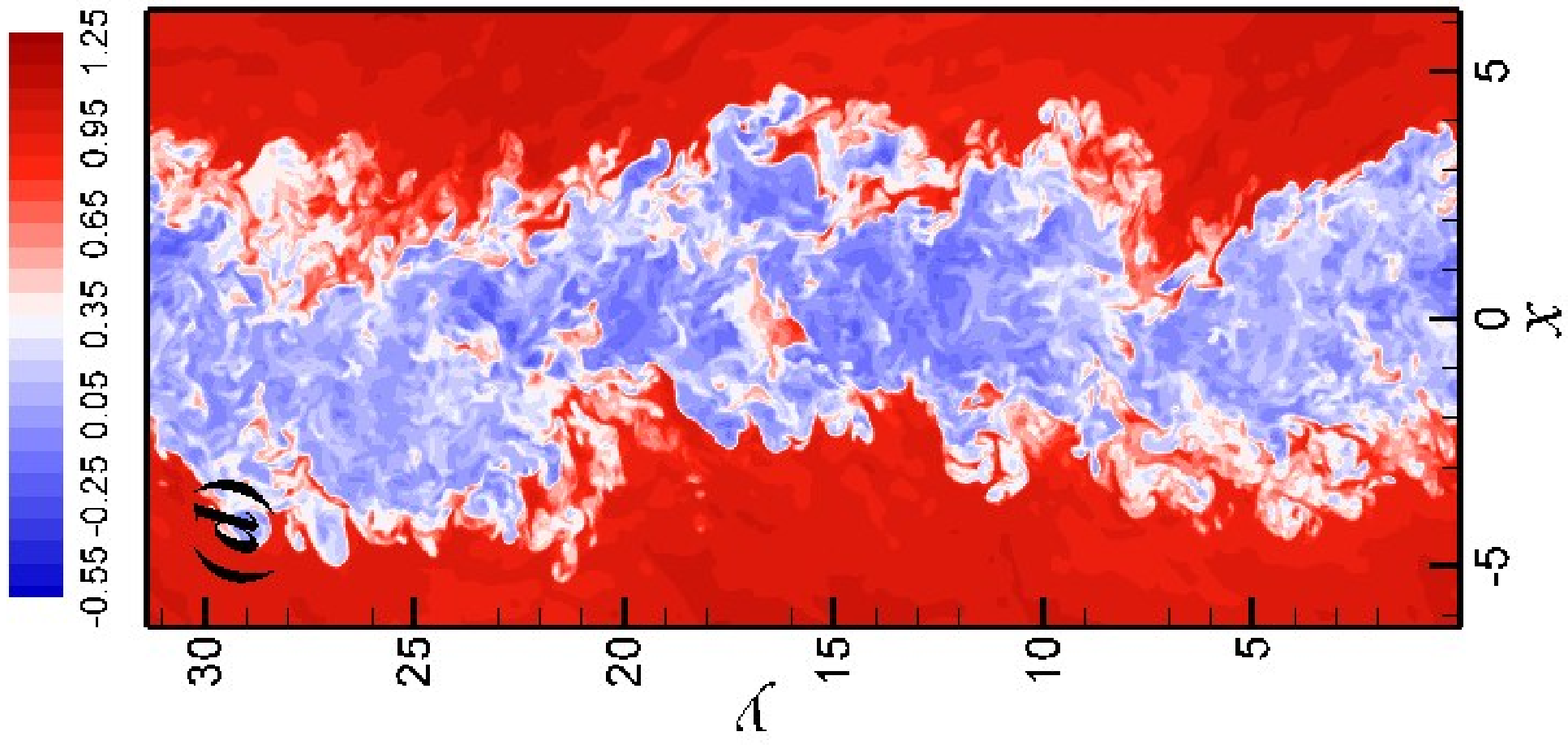}\\ 
\includegraphics[width=0.46\textwidth,angle=-90]{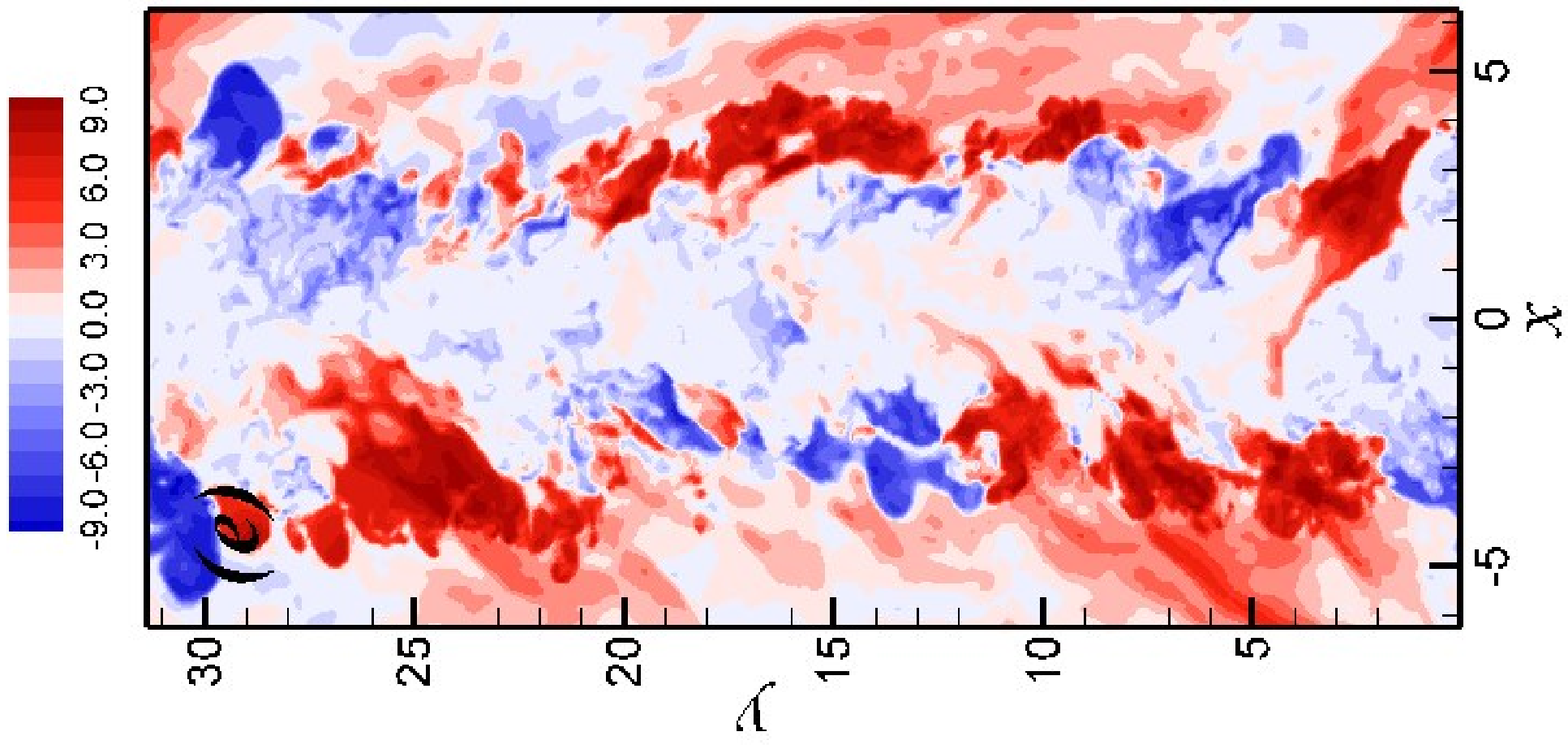} 
\includegraphics[width=0.46\textwidth,angle=-90]{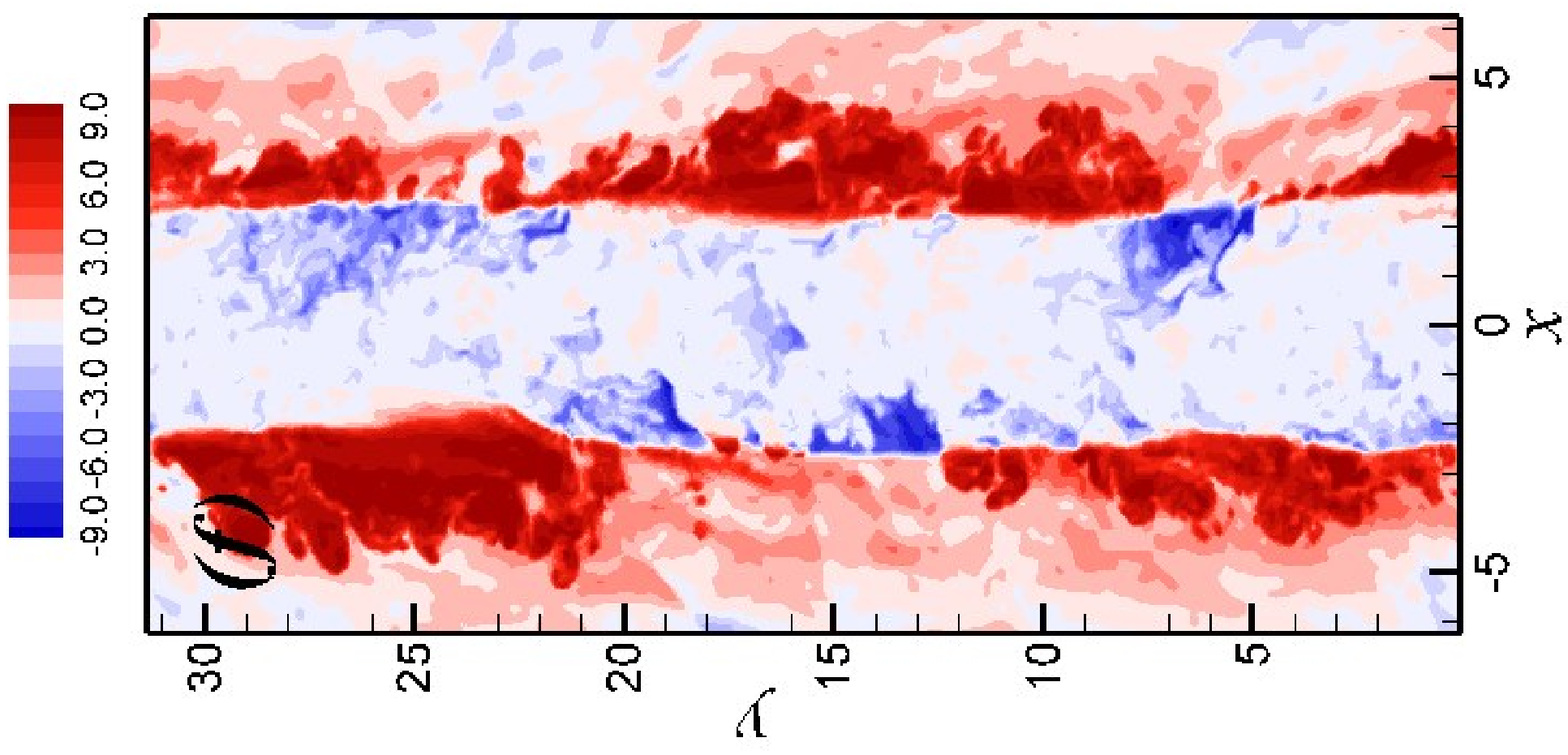} 
\includegraphics[width=0.46\textwidth,angle=-90]{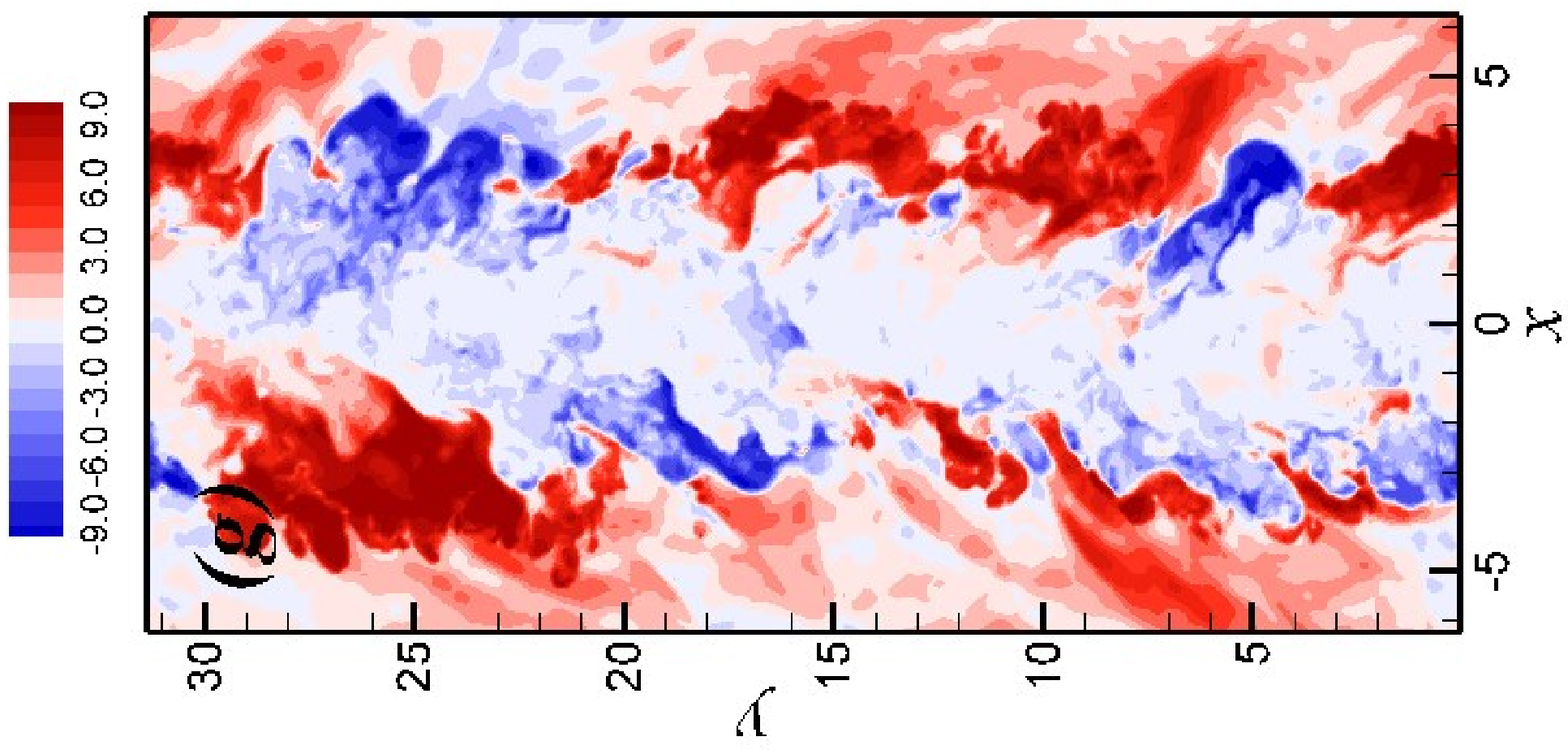} 
\end{tabular}
\caption{Top panel: visualization of the density variation in a longitudinal section at $t=32$: (a) selective LES, (b) standard LES, (c) low resolution pseudo-DNS, (d) higher resolution pseudo-DNS. The figures show the contour levels of $\log_{10} (\rho/\rho_{0j})$, the mean flow is from bottom to top.
Bottom panel: visualization of the local difference between the lower resolution simulations  and the higher resolution pseudo-DNS density fields at $t/\tau=32$: (e) selective LES, (f) standard LES, (g) low resolution pseudo-DNS. The figures show the contour levels of $(\rho_{LES}-\rho_{DNS})/\rho_{0j}$. For a quantification of the average differences along the radial direction see figure 8, panels d and e.}
\label{vis.rho}
\end{figure*}

\begin{figure*}
\centering
\begin{tabular}{c}
\includegraphics[width=0.46\textwidth,angle=-90]{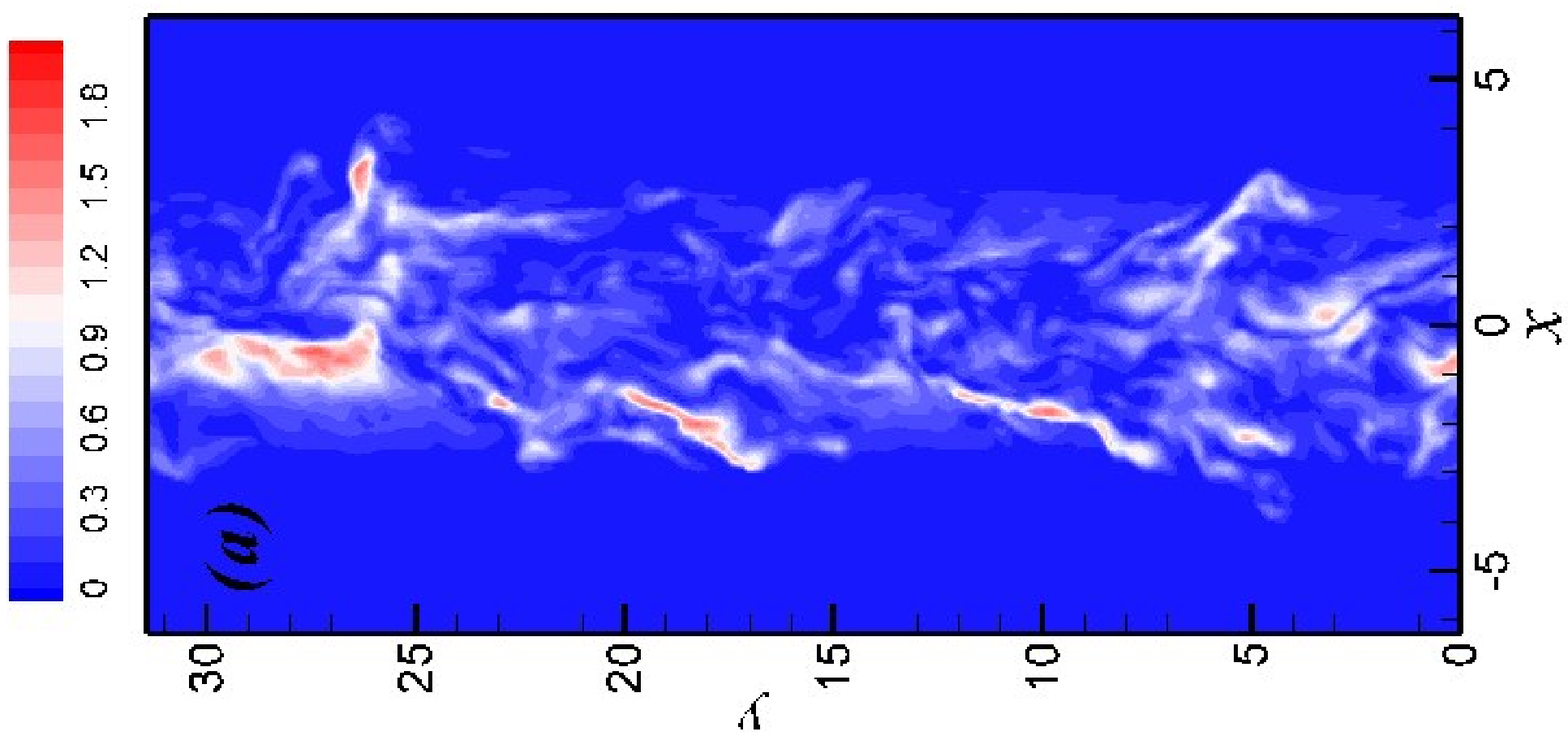}
\includegraphics[width=0.46\textwidth,angle=-90]{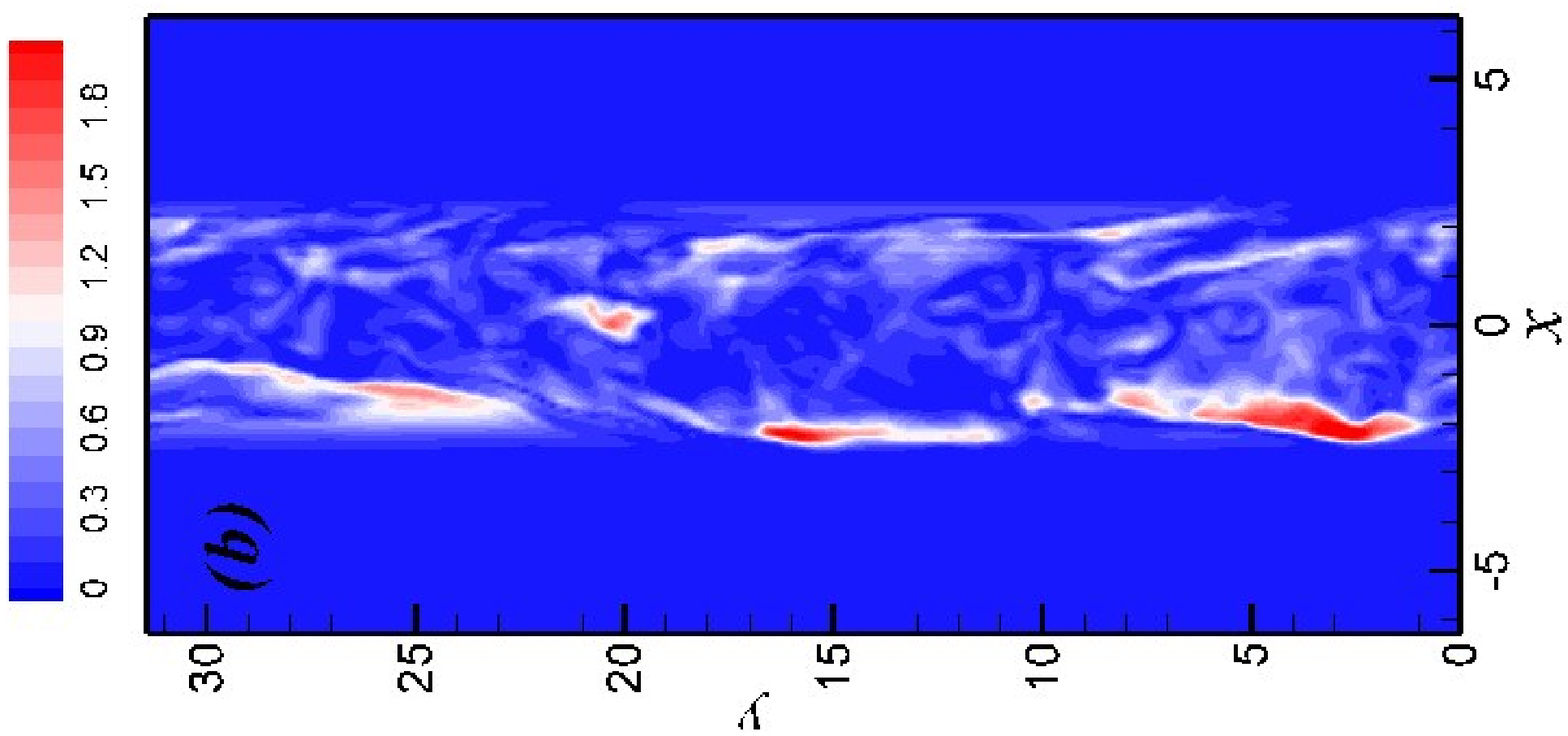}
\includegraphics[width=0.46\textwidth,angle=-90]{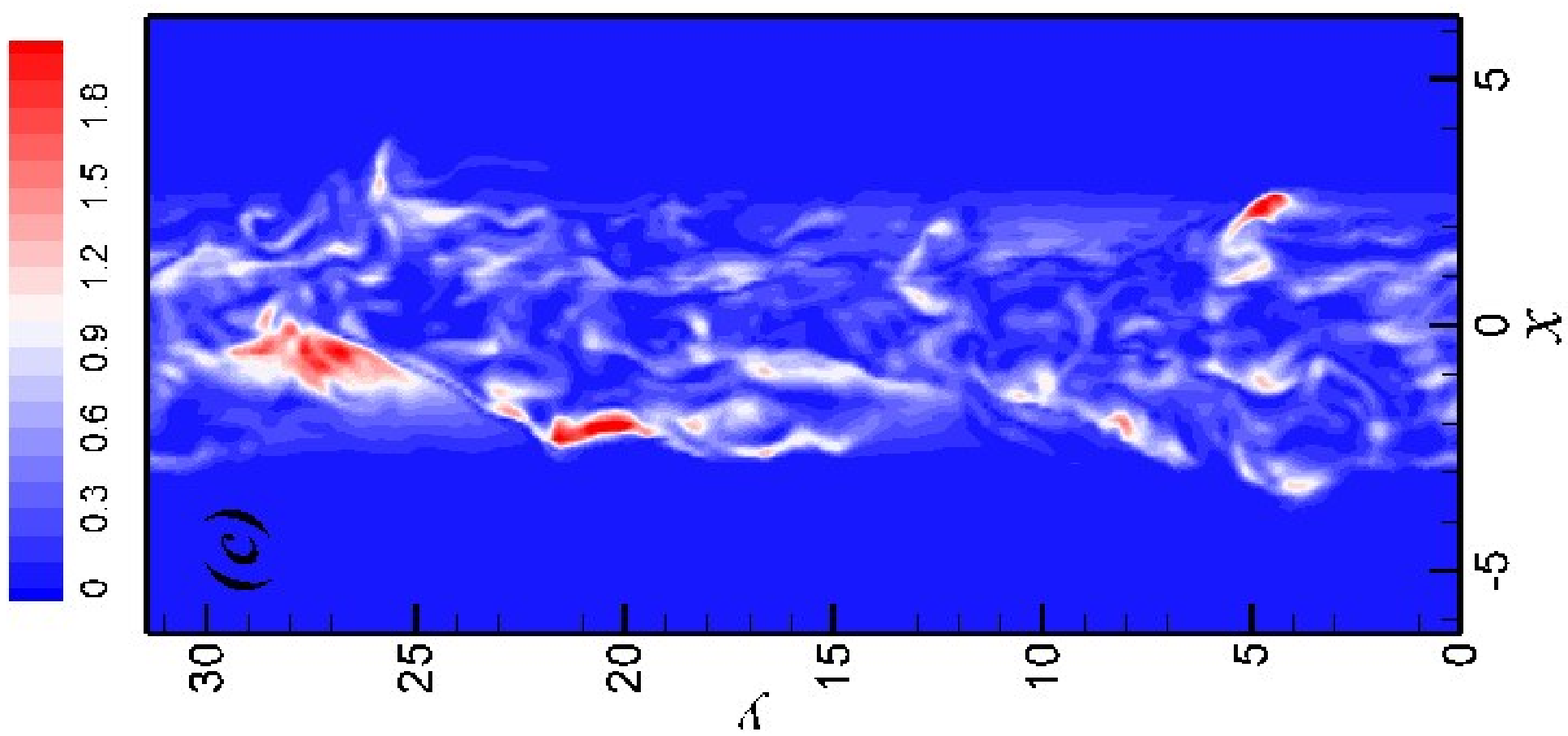}
\includegraphics[width=0.46\textwidth,angle=-90]{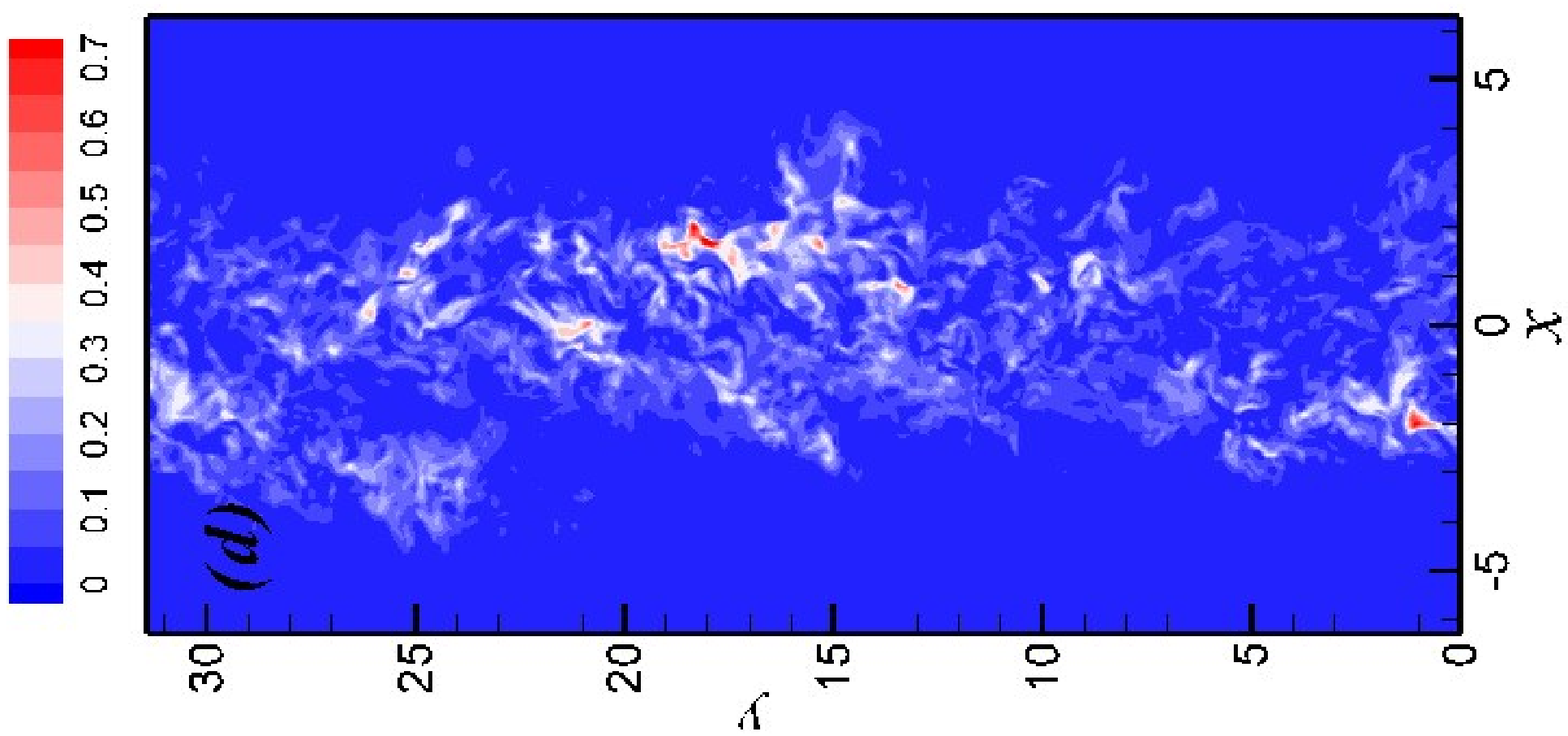}\\
\includegraphics[width=0.46\textwidth,angle=-90]{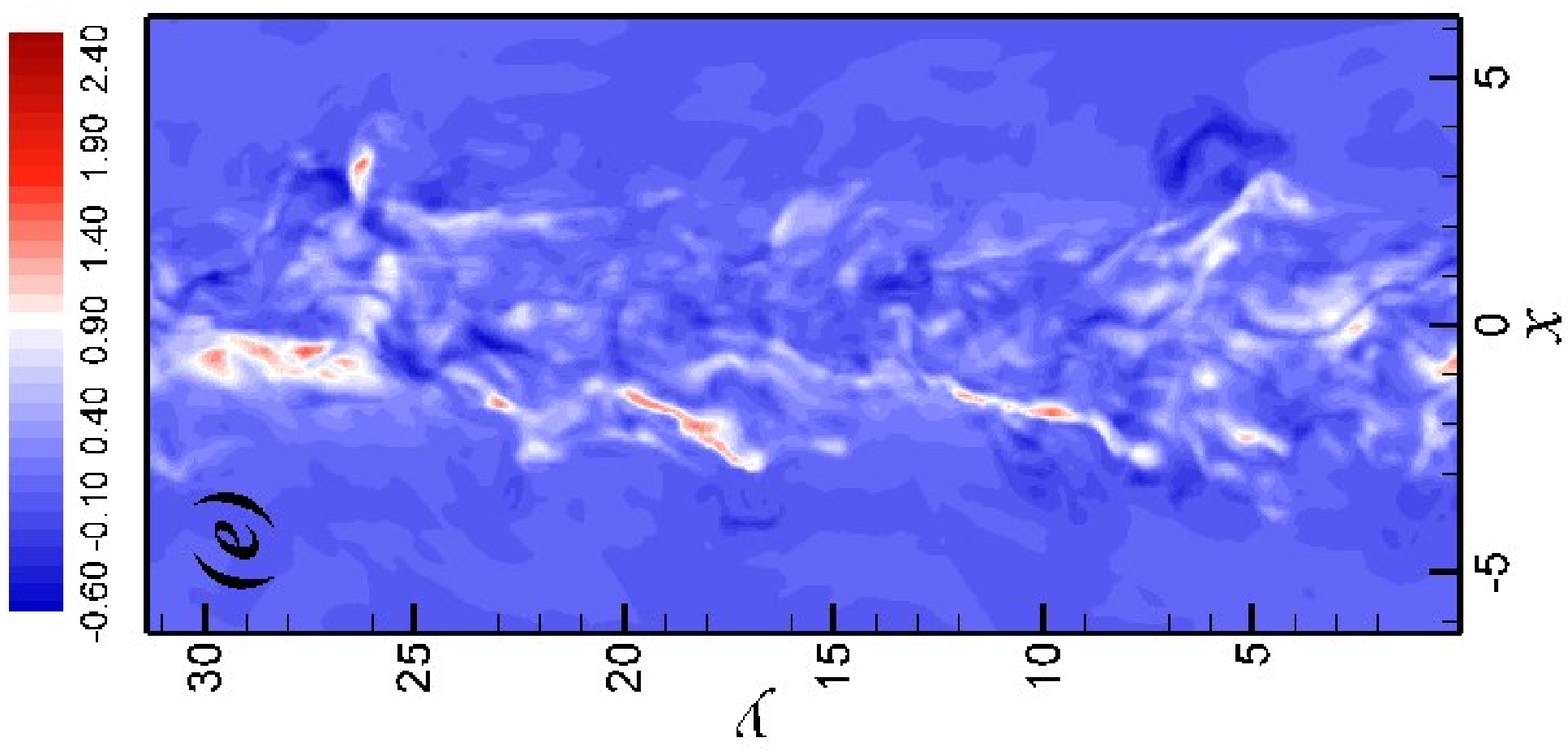}
\includegraphics[width=0.46\textwidth,angle=-90]{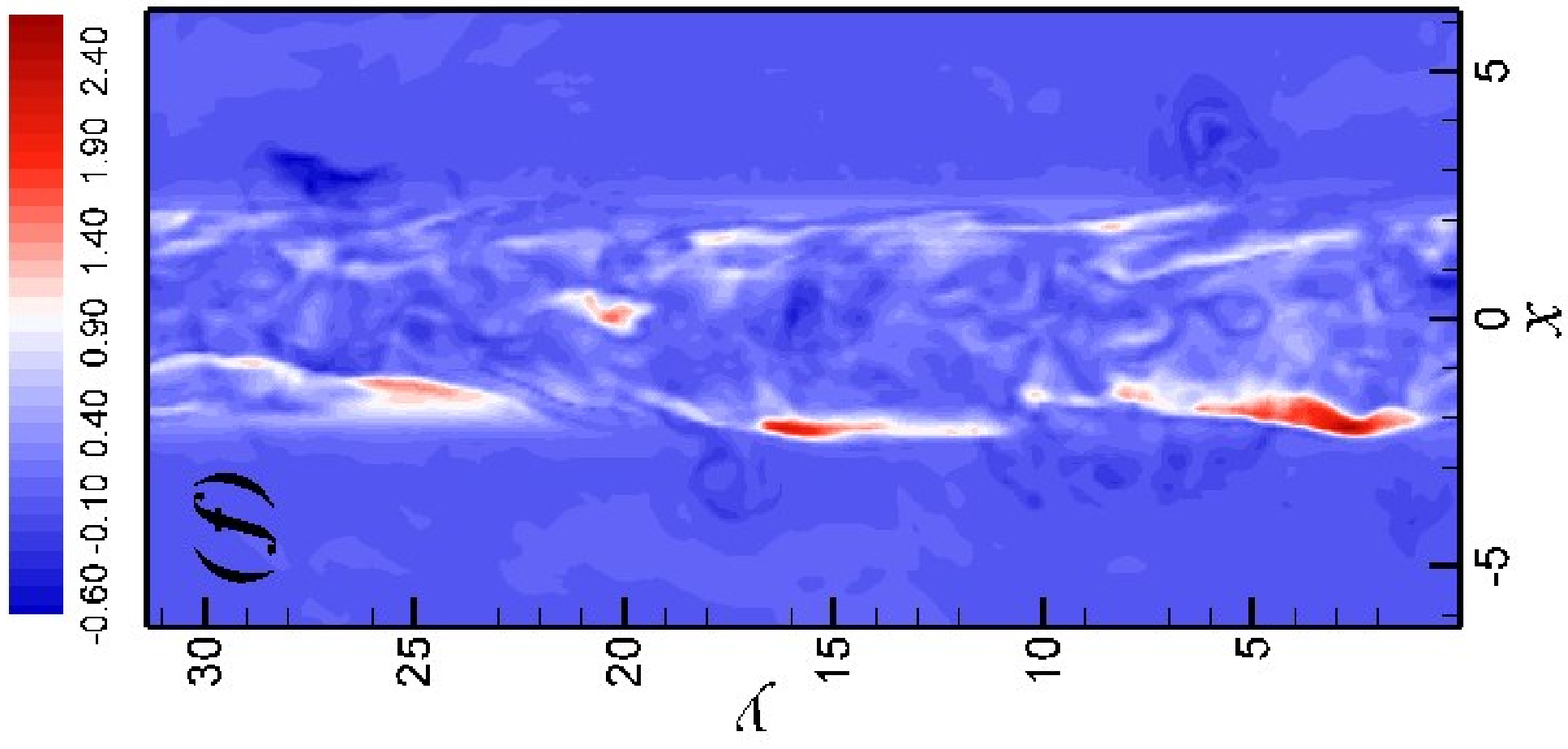}
\includegraphics[width=0.46\textwidth,angle=-90]{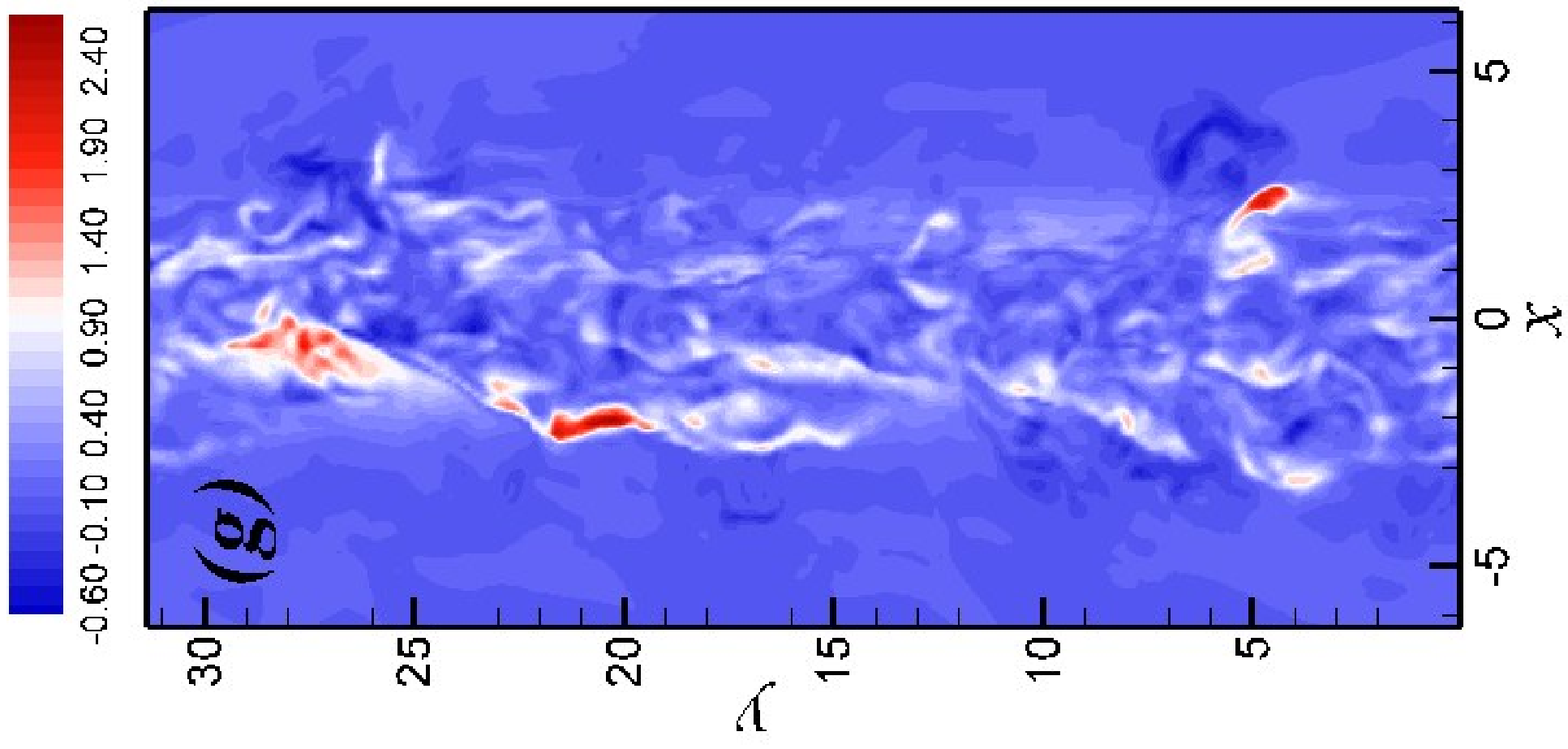}
\end{tabular}
\caption{Top panel: kinetic energy per unit mass of the fluctuating velocity field in a longitudinal section at $t=28$: (a) selective LES, (b) standard LES, (c) low resolution pseudo-DNS, (d) higher resolution pseudo-DNS. The figures show the contour levels of $E=(\tilde{u_x'^2}+\tilde{u_y'^2}+\tilde{u_z'^2})/2$, the mean flow is from bottom to top.
Bottom panel: local difference between the lower resolution simulations and the higher resolution pseudo-DNS at $t/\tau=28$: (e) selective LES, (f) standard LES, (g) low resolution pseudo-DNS. The figures show the contour levels of $(E_{LES}-E_{DNS})$. Note that in this panel the zero is not central with respect to the legend and is represented by a light blue nuance. This is due to the fact that low resolution simulations retain more energy with respect to the high resolution one.}
\label{vis.ene}
\end{figure*}

We have simulated the temporal evolution of a three dimensional jet in a parallelepiped domain with periodicity conditions along the longitudinal direction. The flow is governed by the ideal fluid equations (mono-atomic gas flow) for mass, momentum, and energy conservation. The beam is considered thermally confined by the external medium, and  the initial  pressure is  set uniform in the entire domain.  In the astrophysical context, this  formulation is usually considered to approximate the temporal hydrodynamic evolution inside a spatial window of interstellar jets, which are highly compressible collimated jets characterized by Reynolds numbers of the order $10^{10-15}$. See for example, the Herbig-Haro jets HH24, HH34 and HH47 \cite{be99,rb01}.
We do not consider the effect of the radiative cooling, which can change the jet dynamics substantially (see, e.g. \cite{hardee00,rossi97}).
The transient evolution  includes basically two principal mechanism, the growth and evolution of internal shocks and the dynamics of the mixing process originated by the nonlinear development of the Kelvin Helmholtz instability. The analysis  is carried out through hydro-dynamical simulations by considering only a fraction of the beam which is far from its base and head. Due to the use of the periodic boundary conditions, the jet material is continually processed by the earlier evolution because of the multiple transits though the computational domain.
In this way, the focus is put on the instability evolution and on the interaction between the jet and the external medium, rather than an analysis of the global evolution of the jet.

It is known that the numerical solution of a system of ideal conservation laws (such as the Euler equations) actually produces the equivalent solution of another modified system with additional diffusion terms. With the discretizations used in this study it possible to estimate {\it a posteriori} that the numerical viscosity implies an actual Reynolds number of about $10^3$. In such a situation it is clear that the addition into the governing equations of the diffusive-dissipative terms relevant to a Reynolds number in the range $10^{10-15}$  would be meaningless. The formulation used is thus the following:
\begin{eqnarray}
\!\!\!\!\!
 \pdd{\rho}{t} + \pdd{}{x_i}\Big(\rho u_i\Big) & = & 0
 \label{eq.cont}\\
\!\!\!\!\!\!\!\!\!\!\!\!\!\!\!
 \pdd{(\rho u_k)}{t} + \pdd{}{x_i}\Big(\rho u_iu_k + p\delta_{ik}\Big) &=&
    \pdd{}{x_i}H\left(f_{\rm LES} - t_\omega\right)\tau^{SGS}_{ik}\nonumber\\
  &&
  \label{eq.qmoto}\\
\!\!\!\!\!
  \pdd{E}{t} + \pdd{}{x_i}\Big[(E + p)u_i\Big] &=&
   \pdd{}{x_i}H(f_{\rm LES}-t_\omega)q_i^{SGS}\nonumber\\
 &&
   \label{eq.ene}
\end{eqnarray}
where the field variables $p$, $\rho$ and $u_i$ and $E$ are the filtered pressure, density, velocity, and total energy respectively. The ratio of specific heats $\gamma$ is equal to $5/3$. Here $\tau^{SGS}_{ik}$ and $q_i^{SGS}$ are the subgrid stress tensor and total enthalpy flow, respectively. Function $H(\cdot)$ is the Heaviside step function, thus the subgrid scale fluxes are applied only in the regions where $f > t_\omega$. The threshold $t_\omega$ is here taken equal to 0.4, which is the value for which the maximum difference between the probability density function $p(f > t_{\omega})$  between the filtered and unfiltered turbulence was observed \cite{tim07}. Sensor $f$, as defined in (\ref{f-DNSfluctuation}), does not depend on the subgrid model used and on the kind of discretization used to actually solve the filtered transport equations. In principle, it can be coupled with any subgrid model and any numerical scheme. We have chosen to implement the standard Smagorinsky model as subgrid model,
$$
\tau^{SGS}_{ij} + \frac{1}{3} \tau^{SGS}_{kk}= \rho\nu_\delta S_{ij},\;\;\nu_\delta=(C_s\delta)^2 \mid S\mid
$$
$$
q_{i}^{SGS} = \rho \frac{\nu_\delta}{Pr_t} \frac{\partial}{\partial x_i} E
$$
where $S_{ij}$ is the rate of strain tensor and $\mid S\mid$ its norm. Constant $C_s$ has been set equal to 0.1, which is the standard value used in the LES of shear flows, and $Pr_t$, the turbulent Prandtl number, is taken equal to 1.
The initial flow configuration is an axially symmetric cylindrical jet in a parallelepiped domain,
described by a Cartesian coordinate system $(x,y,z)$. The initial jet velocity is along the $y$-direction; its
symmetry axis is defined by $(x=0,\,\,z=0)$.
The interface between the jet and the surrounding ambient medium is described by a smooth velocity and density transition in order to avoid the spurious oscillations that can be introduced by a sharp discontinuity.
The longitudinal velocity  profile is thus initialized as
$$
\overline{u}(r)=\frac{U_0}{\cosh(r/a)^m}
$$
where $r^2=x^2+z^2$ is the distance from the jet axis, $R$ is the jet radius and $U_0$ the jet velocity. $m$ is a smoothing parameter which has been set equal to 4. The same smoothing has been used for the initial density distribution,
$$
\overline{\rho}(r)= \rho_{0j} \left(\nu - \frac{\nu-1}{\cosh(r/a)^m}\right)
$$
where $\rho_{0j}$ is the initial density inside the jet ambient and $\nu$ is the ratio between the ambient  density at infinity to that of on the jet axis.
A value of $\nu$ larger than one implies that the jet is lighter than the external medium.
The mean pressure is set to a uniform value $p_0$, that is, we are considering a situation where there is initially a pressure equilibrium between the jet and the surrounding environment.
This initial mean profile is perturbed at $t=0$ by adding longitudinal disturbances on the transversal velocity components whose amplitude is 5\%\ of the jet velocity and whose wavenumber is up to eight times the fundamental wavenumber $2\pi/L_y$,
$$
u_x(x,y,z)= \frac{0.05}{\cosh(r/a)^m}U_0\sum_{n=0}^{8} \sin\left(n\frac{2\pi}{L_y}y+\varphi_n\right)
$$
$$
u_z(x,y,z)= \frac{0.05}{\cosh(r/a)^m}U_0\sum_{n=0}^{8} \cos\left(n\frac{2\pi}{L_y}y+\varphi_n\right)
$$
with $\varphi_n$ random phase shifts, so that even the perturbation with the shortest wavelength is, initially, fully resolved.
The integration domain is $-L_x\le x\le L_x$, $0\le y\le L_y$ and $-L_x\le z\le L_x$, with $L_x=2\pi R$ and $L_y= 10\pi R$. We have used periodic boundary conditions in the longitudinal $y$ direction, while free flow conditions are used in the lateral directions. A scheme of the initial flow configuration used in the simulations is shown in figure \ref{fig.schema}.

\begin{figure}
\centering
\psfrag{W}{\large$\overline{\omega'_i\omega'_i}$}
\includegraphics[width=0.89\columnwidth]{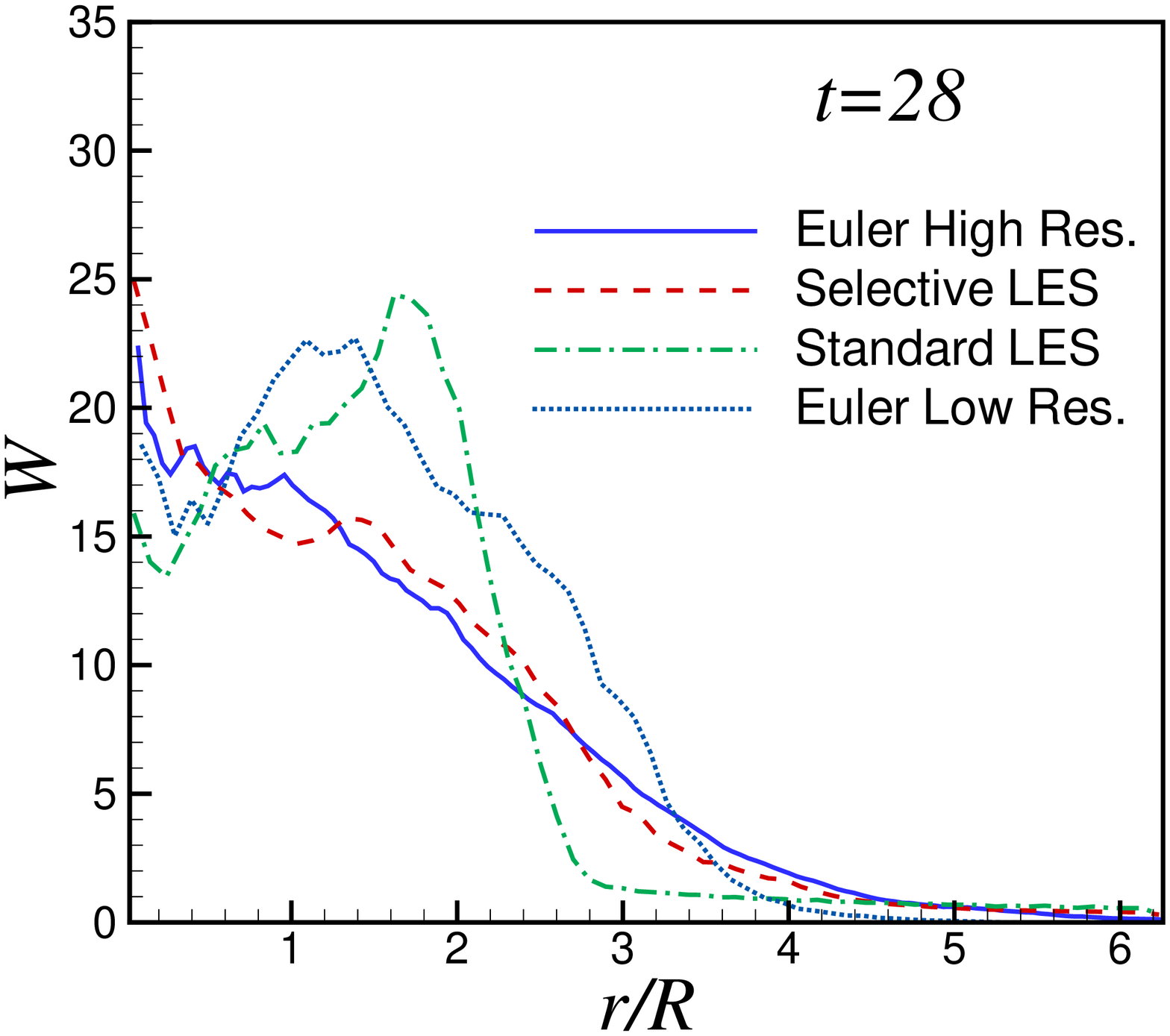}\\  
\includegraphics[width=0.89\columnwidth]{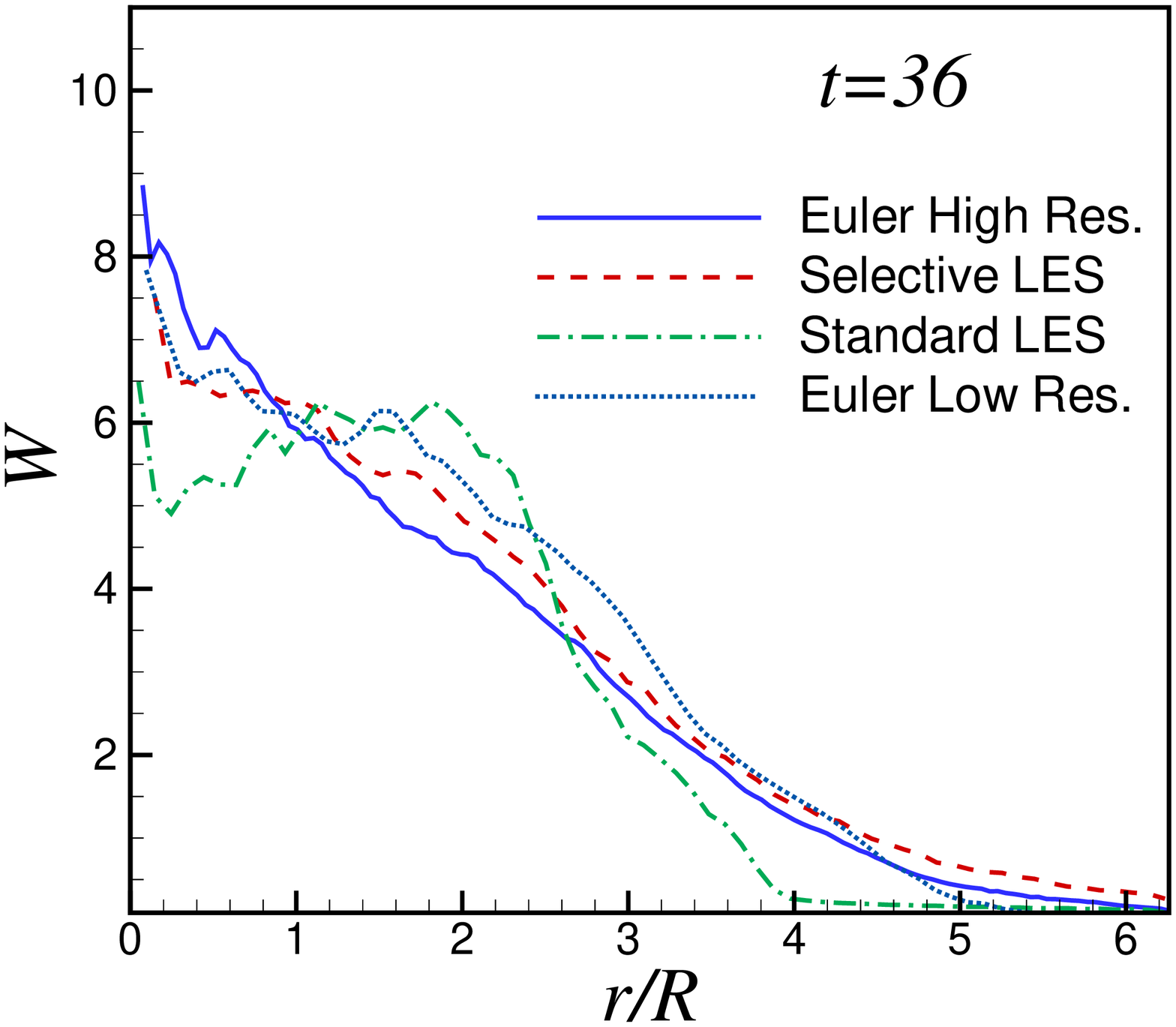}
\vskip -1.0mm
\caption{Radial distribution of the enstrophy $\overline{\omega'_i\omega'_i}$ as function of the distance $r$ from the axis of the jet. All averages have been computed as space averages on cylinders at constant $r=2$.}
\label{fig.enstrofia}
\end{figure}

In the following, all data have been made dimensionless by expressing lengths in units of the initial jet radius $R$, times in units of the sound crossing time of the radius $R/c_{\rm 0}$, where $c_{\rm 0}=\sqrt{\gamma p_0/\rho_{0j}}$ is the reference sound velocity of the initial conditions, velocities in units of $c_{\rm 0}$ (thus dimensionless velocities coincide with the initial Mach number), densities in units of $\rho_{0j}$ and pressures in units of $p_0$.

Equations (\ref{eq.cont}-\ref{eq.ene}) have been solved, in Cartesian geometry, using an extension of the PLUTO code \cite{pluto}, which is a  Godunov-type code that supplies a series of high-resolution shock-capturing schemes \cite{cw84} that are particularly suitable for the present application, because of their low numerical dissipation. In fact, as pointed out by \cite{d99}, a high numerical viscosity can overwhelms the subgrid-scale terms effects. The code has been extended by adding the subgrid fluxes and the computation of the functional $f$ which allows to perform the selective large-eddy simulation. For this application, a third order accurate in space and second order in time Piecewise-Parabolic-Method (PPM) has been chosen. 


We have performed three simulations of a jet with an initial Mach number equal to 5 and a density ratio $\nu$ equal to 10. The density ratio is an important parameter in such flow configuration, as it has been shown that it has a strong influence on the temporal evolution and on the flow entrainment as it has been shown by numerical simulations and laboratory experiments \cite{njp2011}.
The selective LES of the jet has been carried out on a $320\times 128^2$ uniform grid. A uniform grids avoids the need to cope with the non-commutation terms in the governing equations (see, e.g. \cite{it03}). Moreover, three additional simulations were performed for comparison: a standard non selective LES where the subgrid model was introduced in the whole domain, which is obtained by forcing $H\equiv1$ in equations (\ref{eq.cont}-\ref{eq.ene}), and two Euler simulations, which formally can be obtained by putting $H\equiv0$, one with the same resolution of the Large Eddy Simulations and one which uses a finer grid ($640\times 256^2$).

\section{Results}

\begin{figure*}
\centering
\begin{tabular}{cc}
\includegraphics[width=0.38\textwidth]{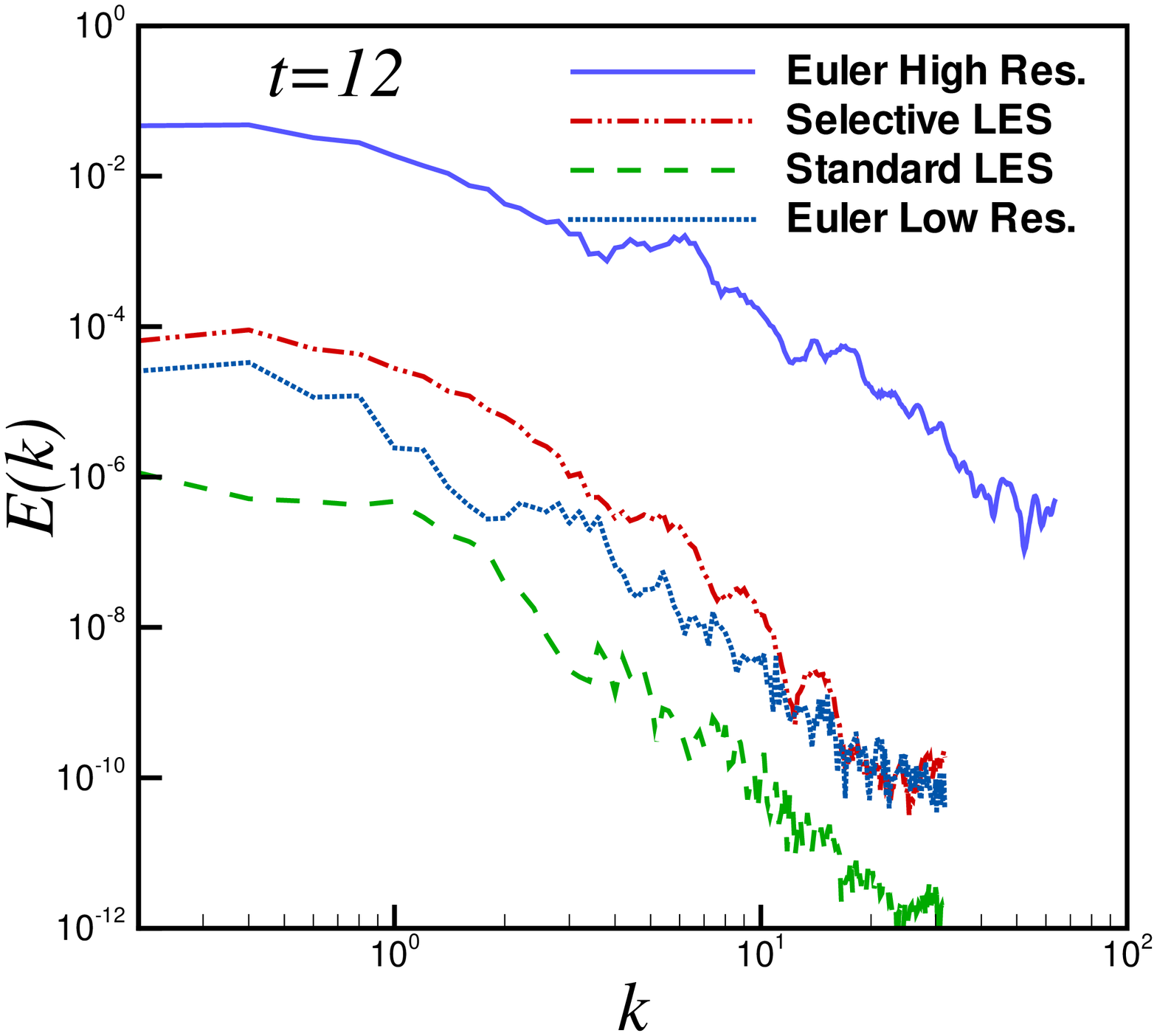}&
\includegraphics[width=0.38\textwidth]{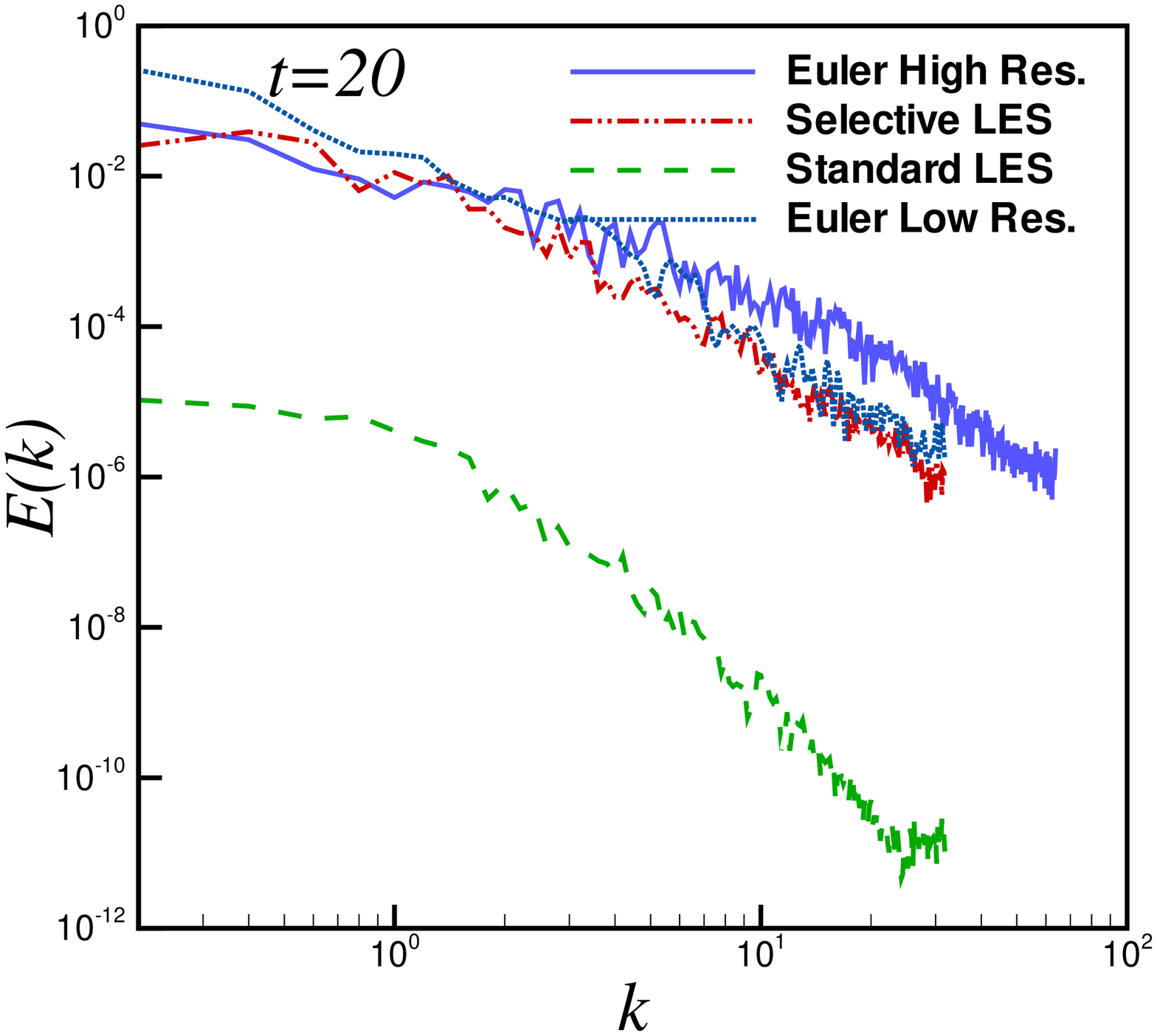}\\
\includegraphics[width=0.38\textwidth]{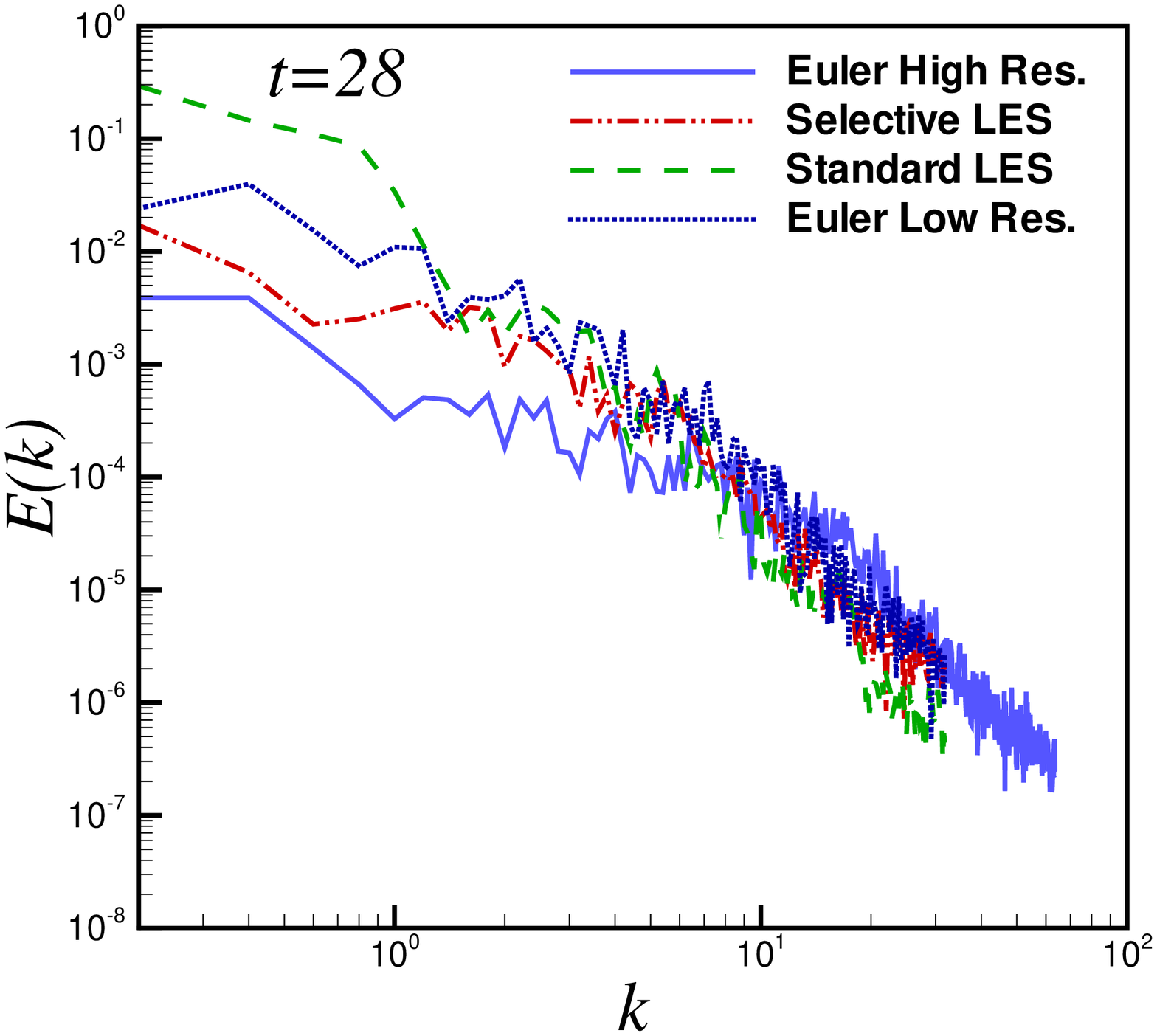}&
\includegraphics[width=0.38\textwidth]{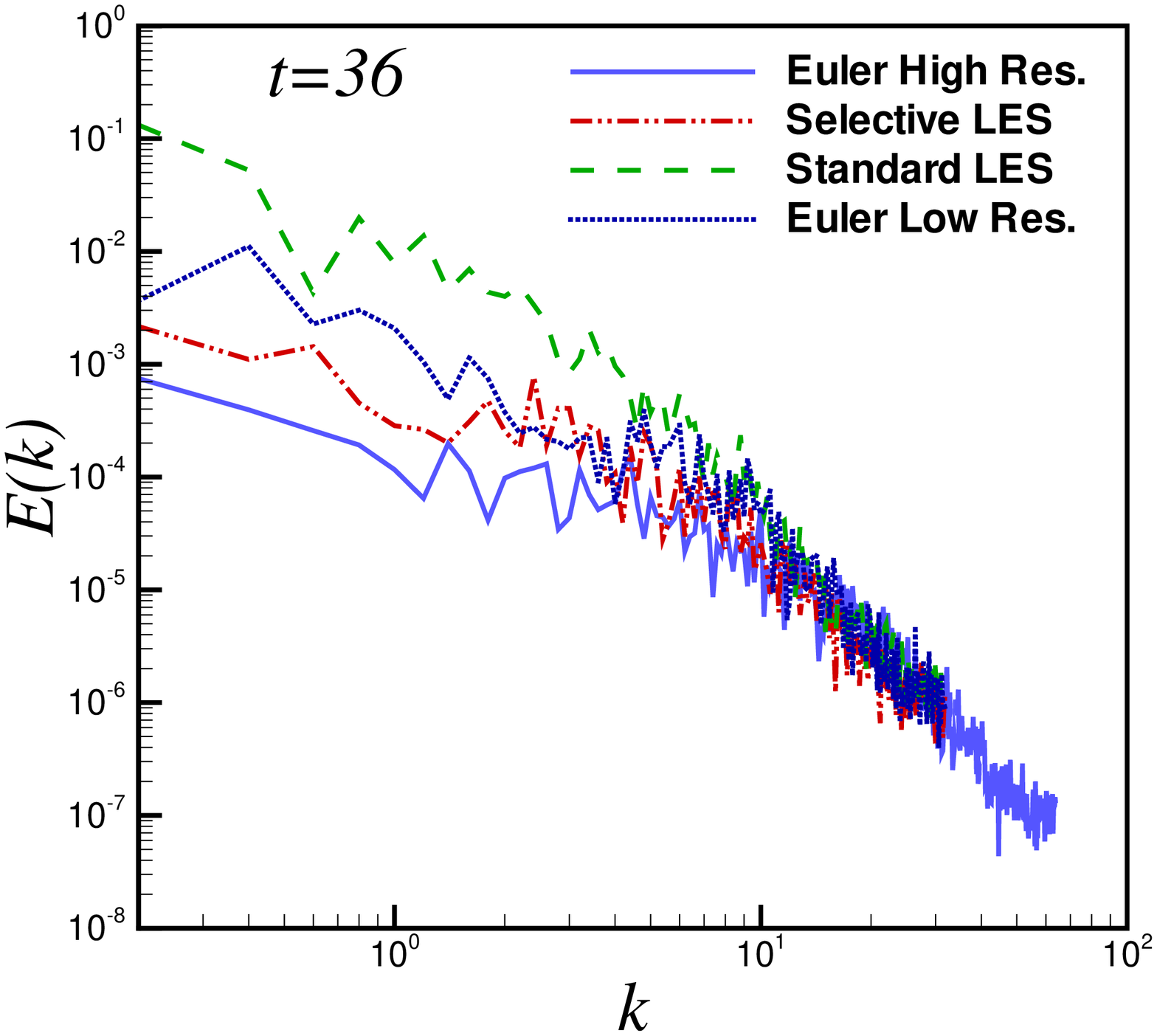}
\end{tabular}
\caption{Longitudinal one dimensional spectra of the turbulent kinetic energy obtained by considering all data on the cylindrical surface at $r=2$, computed as the Fourier transform of the two-point correlations of the fluctuating kinetic energy $\rho u_i u_i/2$.}
\label{fig.spettri}
\end{figure*}

\begin{figure*}
\centering

\begin{tabular}{cc}
\includegraphics[width=0.38\textwidth]{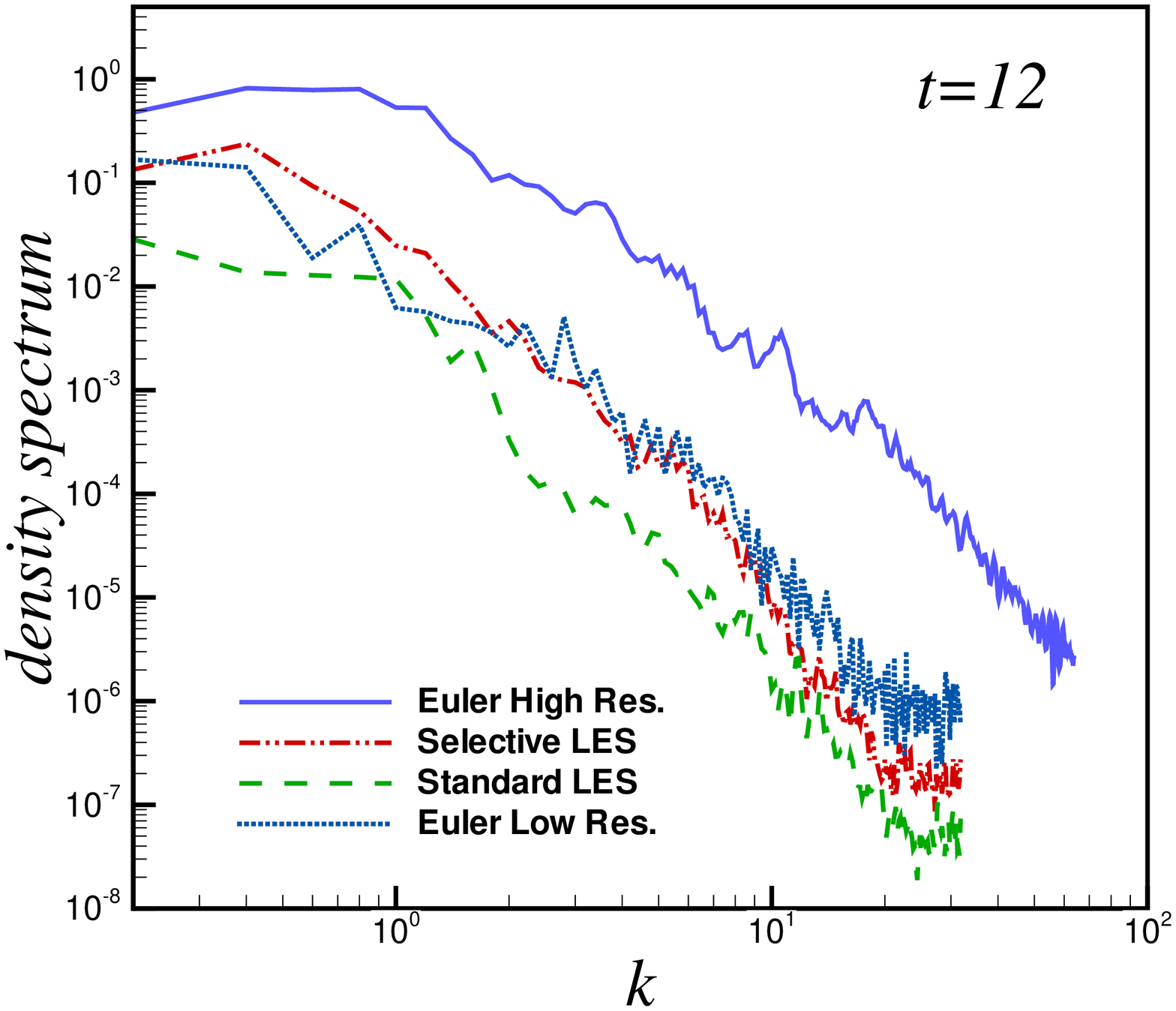}&
\includegraphics[width=0.38\textwidth]{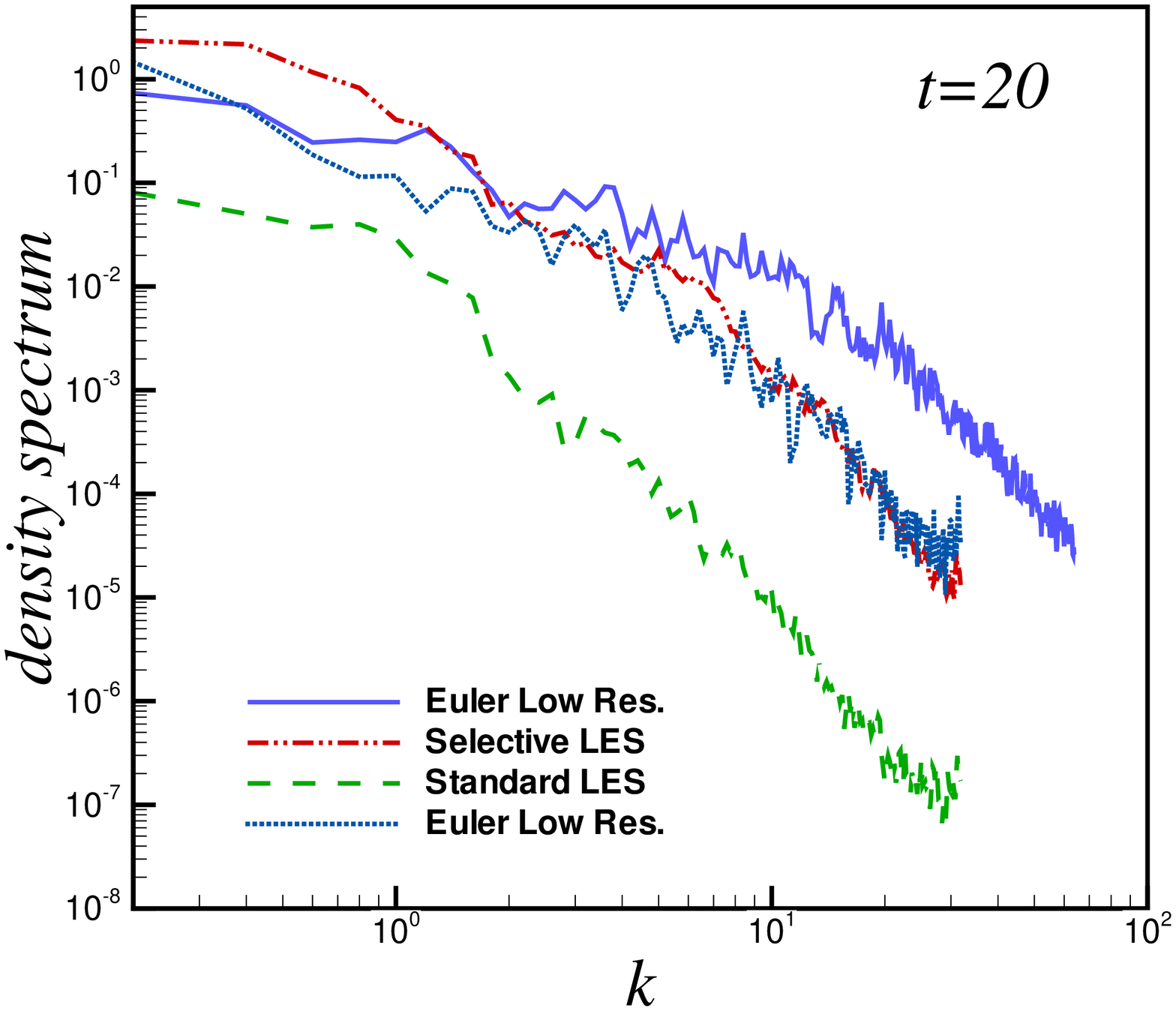}\\
\includegraphics[width=0.38\textwidth]{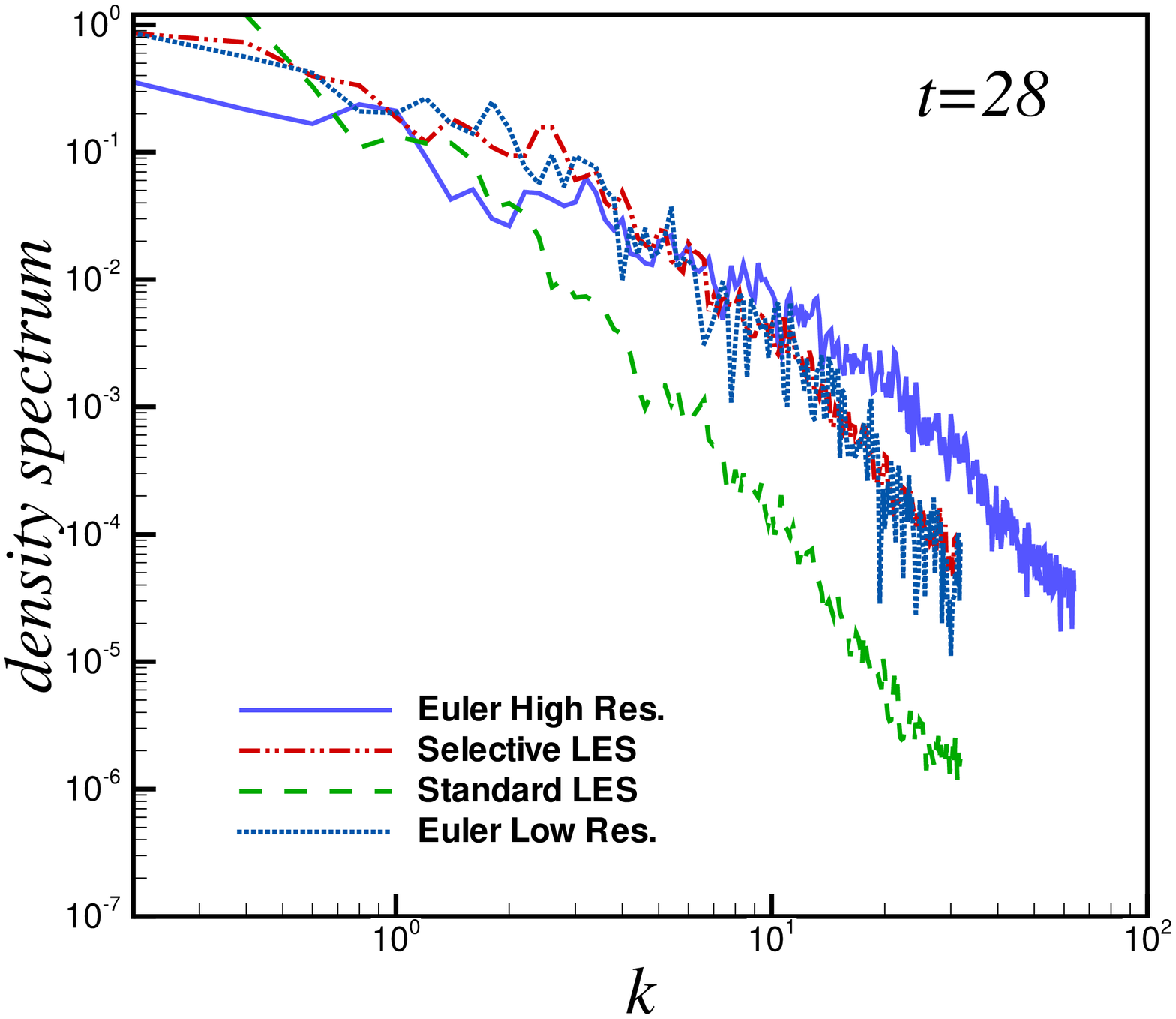}&
\includegraphics[width=0.38\textwidth]{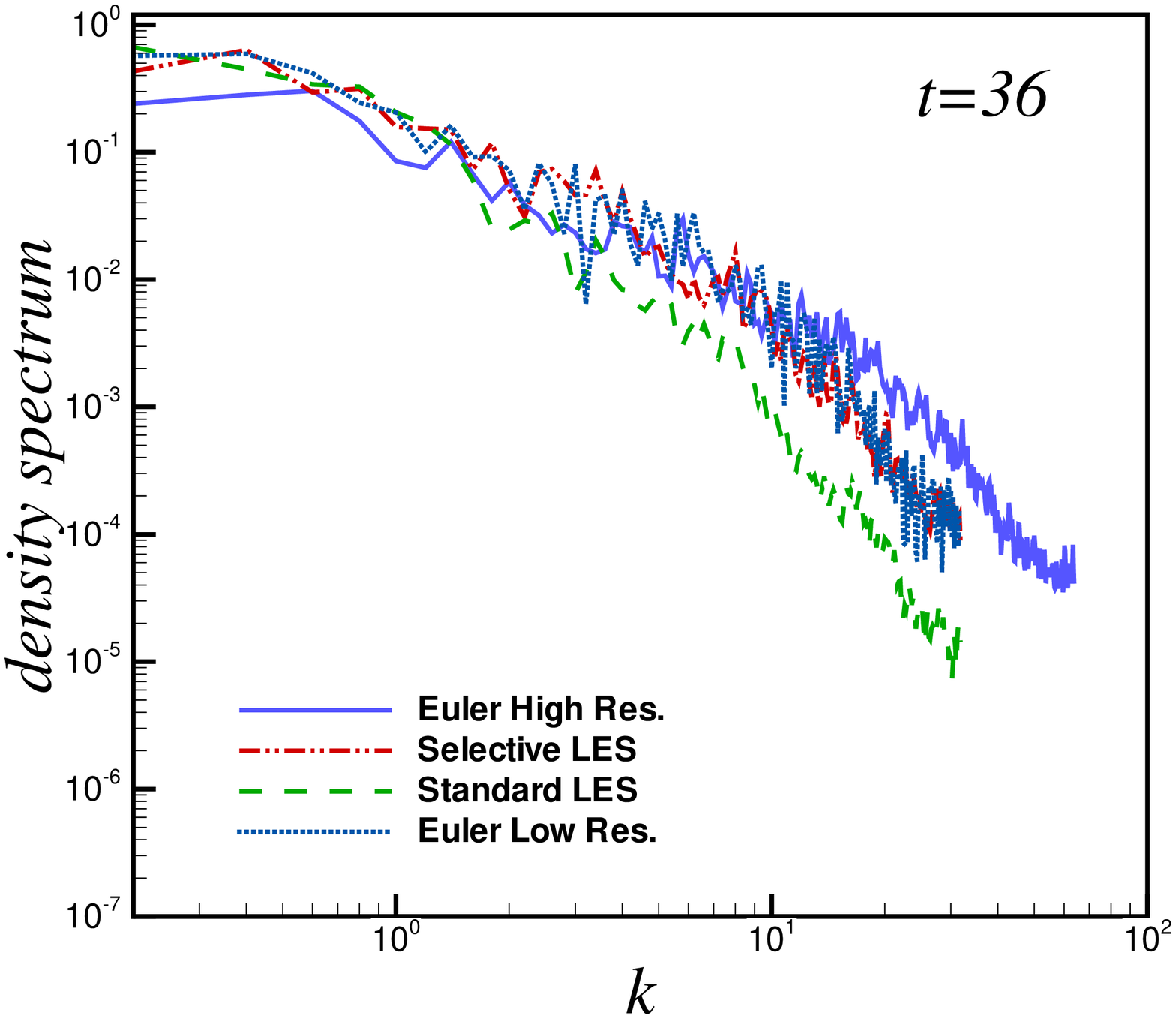}
\end{tabular}

\vskip -1.0mm
\caption{Longitudinal one dimensional spectra of the density fluctuations obtained by considering all data  at a distance from the jet axis equal to $r=2$.}
\label{fig.spettri-rho}
\end{figure*}

In this study we follow the temporal evolution of a portion of a highly compressible jet with an initial Mach number equal to 5.
Our physical  system is that of an ideal highly compressible jet flow represented by an Initial Value Problem formulation, where the initial  energy is not forced to be  constant in time and the initially smooth distribution of momentum, density and thermal energy per unit volume are perturbed by a set of eight longitudinal velocity disturbances  randomly out of phase. The largest wavelength is $10\pi$ times the initial radius of the jet and the amplitude is $5 \%$
of the initial mean velocity.
The model that we use considers the fluid as ideal and focus on the hydrodynamics aspects of hypersonic jets as viewed in the astrophysical context. In synthesis, through momentum conservation,  due to the interaction with the ambient fluid and the associated entrainment process and  the loss associated to the formation of shocks and related  acoustic emission, the velocity  inside the jet decreases, while the width  and the mass flow grow. 
It should be noted that our computational domain is an observation window where we see the  full interaction between the jet and the ambient  up to about 6 times scales, which is the time where the information reaches the computational frontier. Afterwards  the visibility on the interaction is partial and is limited to the central part of the physical system. Anyway,  in the computational domain, the total energy remains constant in the first stage of the evolution. The kinetic energy decay, in fact,  is initially compensated by the thermal energy increase due to the shocks formation and the (numerical) dissipation activated by the development of the small scales.  Afterwards,   the kinetic energy of the fluctuation decays as well as   the related  dissipation. The total energy in the computational domain decays for $t > 30$.

As it is known from previous studies on the subject (e.g.\ \cite{2000,bm98,hardee95,dr98,hardee00,rossi97} for three-dimensional jets and \cite{hardee88, bm94,bodo95} for two dimensional or axisymmetric jets), four main stages can be identified in the temporal evolution of hypersonic jets. In the first phase, the unstable modes introduced by the perturbations grow up in agreement to the linear theory. In the end of this stage, their growth leads to the formation of internal shocks. This stage is followed by a second phase where the jet is globally deformed and shocks are driven in the external medium, thus carrying momentum and energy away from the jet.
At this point a third stage, called mixing stage, occurs: as a consequence of the shock evolution, mixing between the jet and external material begins to occur. The longitudinal momentum, initially concentrated inside the jet radius, is spread over a much larger region by the mixing process. In the fourth final stage the jet can reach a statistically quasi-stationary phase if over-dense with respect to the ambient, otherwise the mean and fluctuating velocity field  decay, a light jet in the long term is in fact going to be essentially destroyed by the interaction with the ambient medium. An overview of the thermal and kinetic energy time evolution of the jets here studied can be found in figure 10.

The onset of these stages depend upon the values of the Mach number and density ratio, and the choice of the initial perturbation amplitudes and phase can change the temporal length of the initial stage, but the global pattern of the jet evolution remain unchanged \cite{bm98, bodo95, hardee88}.

The initial growth of the kinetic energy of the fluctuations is associated to the formation of small scales.
The mixing phase where the flow can be considered turbulent is reached after about 15--20 initial sound crossing times. At this point, the resolution could not be enough to solve all the scales and, consequently, the momentum and energy transport due to presence of sub-grid scales should be introduced: the sub-grid terms must be active in the under-resolved regions.
Figure \ref{getto-prob} shows, in our selective large-eddy simulation, the probability that the sensor $f$ is larger than the threshold, that is, that sub-grid scales are present, at $t=28$. At this stage about 40-60\%\ of the jet is under-resolved and sub-grid terms are applied in such zones. At the same time, the external ambient is still resolved with the LES grid.
In the last stage of the simulation the total energy of the flow inside the computational domain reduces by about 20\%\ due to the reduced dissipation associated to the decay of the fluctuation kinetic energy and to lateral acoustic wave radiation, see again  \cite{bm98}. The mean flow becomes subsonic -- the Mach number becomes about 0.5 at the end of the simulation -- and the kinetic energy of the fluctuations is reduced by about 40\%. In this phase the extension of the under-resolved regions is likewise reduced and the effect of the inclusion of sub-grid terms on the flow dynamics becomes gradually milder.

The effect of the subgrid scale terms can be qualitatively appreciated in the visualizations of the pressure and density fields. A visualization of the pressure field in a longitudinal section at $t=32$ is shown in figure \ref{vis.pr}(a-d) for the four simulations (selective LES, classical LES, low resolution Euler and high resolution Euler simulation). The comparison shows the larger dissipation and the small scale suppression produced by the non selective use of the subgrid model in the whole domain. This is even more evident in the plot of the density field (figure \ref{vis.rho}(a-d)): subgrid terms used in the whole volume of flow tend to delay the mixing of the jet and reduce the spreading of the jet material. Parts (e-g) of figures \ref{vis.pr} and \ref{vis.rho} present the difference between the pressure and densities predicted, at $t=32$, by the lower resolution simulations and the higher resolution pseudo-DNS. The selective LES mainly introduces, in comparison with the higher resolution simulation, a shift on the pressure/density fluctuations in the longitudinal direction, while the standard LES mainly tends to suppress the density fluctuations at the jet border and thus reduces the flow  entrainment.

In figure \ref{vis.ene}, analogue visualizations of the kinetic energy  field in  longitudinal jet sections at $t=28$ are shown. This figure is collateral to figure \ref{fig.energia} which presents the evolution of the mean and fluctuating kinetic energy and of the thermal energy in the high resolution simulation. In the lower panel of figure \ref{vis.ene}, the contour levels of $(E_{LES}-E_{DNS})$ are shown. Here, it should be noted that the zero is not central with respect to the legend and is represented by a light blue nuance. This is due to the fact that low resolution simulations typically retain more energy with respect to the high resolution one. Anyway, the more uniformly light blue is the image the lower is the difference with the high resolution flow simulation. One can notice, that the selective LES performs slightly better than the low resolution Euler simulation and definitively better than the standard LES.

The time evolution of the enstrophy distribution at two time instants far from the initial one is shown in figure \ref{fig.enstrofia} as a function of the distance from the centre of the jet.
While the agreement between the enstrophy distribution obtained with the selective LES simulation  and with the reference high resolution Euler simulation is fair, the non selective simulation  damps out the vorticity magnitude in the centre of the jet and in the outer part, and introduces a spurious accumulation in the intermediate radial region. As a results, the vorticity dynamics is highly modified. The overall effect is a delay in the formation of the turbulent structures, as it is evident when the spectrum of the turbulent kinetic energy is considered.
The selective LES performs better than the low resolution Euler simulation: enstrophy profiles remain closer to the ones obtained from the higher resolution pseudo-DNS.

Figure \ref{fig.spettri} and figure \ref{fig.spettri-rho} show the one dimensional longitudinal kinetic energy and the fluctuating density spectra at $r=2$, that is in the intermittent region between the jet core and the surrounding ambient. All spectra have been computed by averaging on points at the same distance from the jet axis. It can be noted that, in the non selective LES, the growth of fluctuations is much slower and, as a consequence, they have much less energy in the first stage of evolution. Moreover, even when the energy of the fluctuations in the non selective LES reaches levels comparable with those of both the selective LES and the higher resolution run ($t=28$ and 36), there is a significant concentration of energy in the low wave-number region, which becomes even more pronounced in the later stages ($t=36$). This is consistent with the higher level of enstrophy seen in figure \ref{fig.enstrofia} for the non selective LES at a similar distances from the centre of the jet. Thus, we can observe that the selective introduction of the sub-grid model yields spectral distributions of the energy much closer, with respect to the standard LES, to the distribution shown by the high resolution Euler simulation.
The Euler low resolution simulation tends, in the final stage of evolution ($t=28$ and 36), to accumulate more energy in the resolved scales than the selective LES, a sign of the inability of the numerical diffusion alone to properly account the energy transfer toward sub-grid scales.

\begin{figure*}
\centering
\begin{tabular}{cc}
\includegraphics[width=0.38\textwidth]{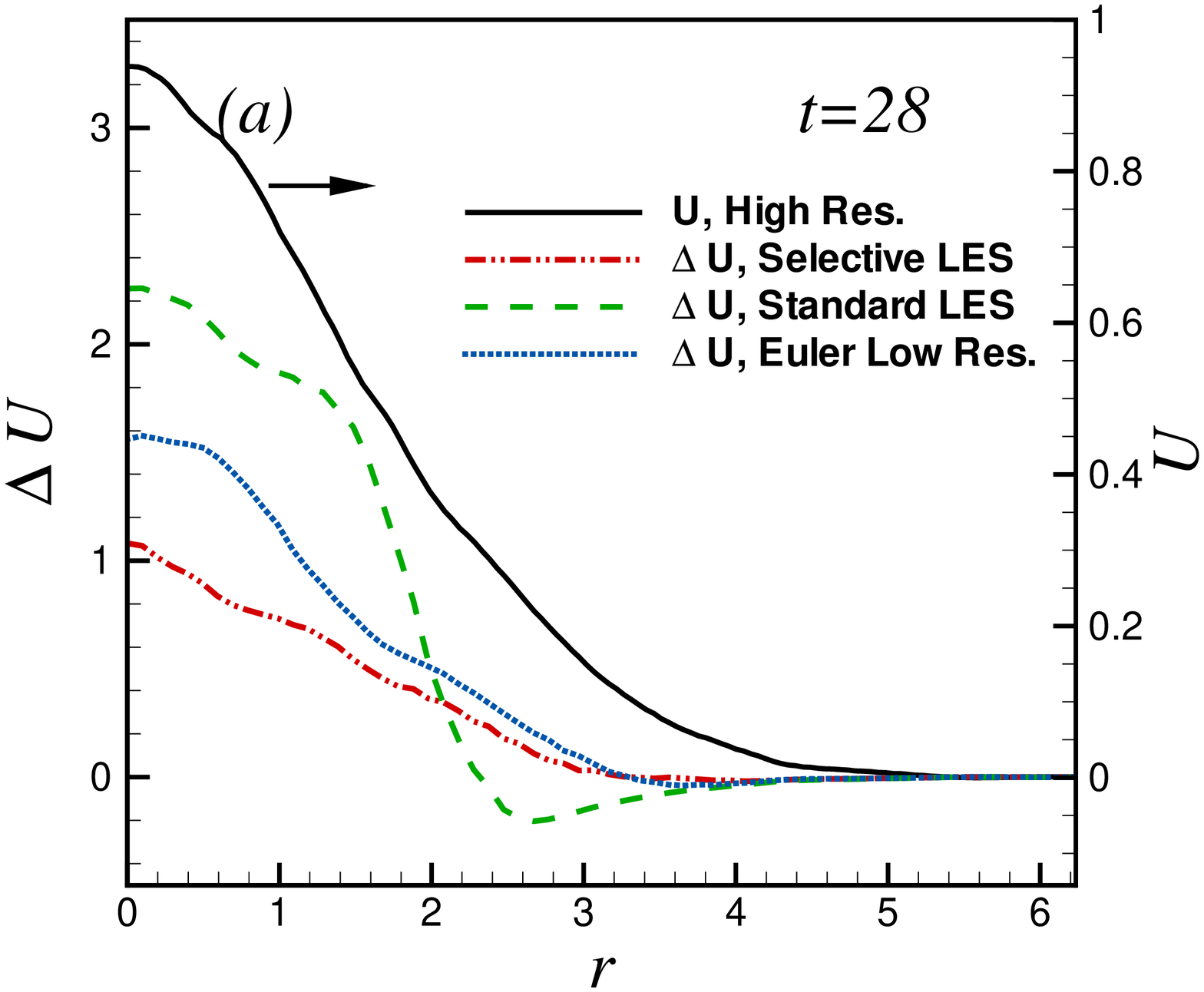}&
\includegraphics[width=0.38\textwidth]{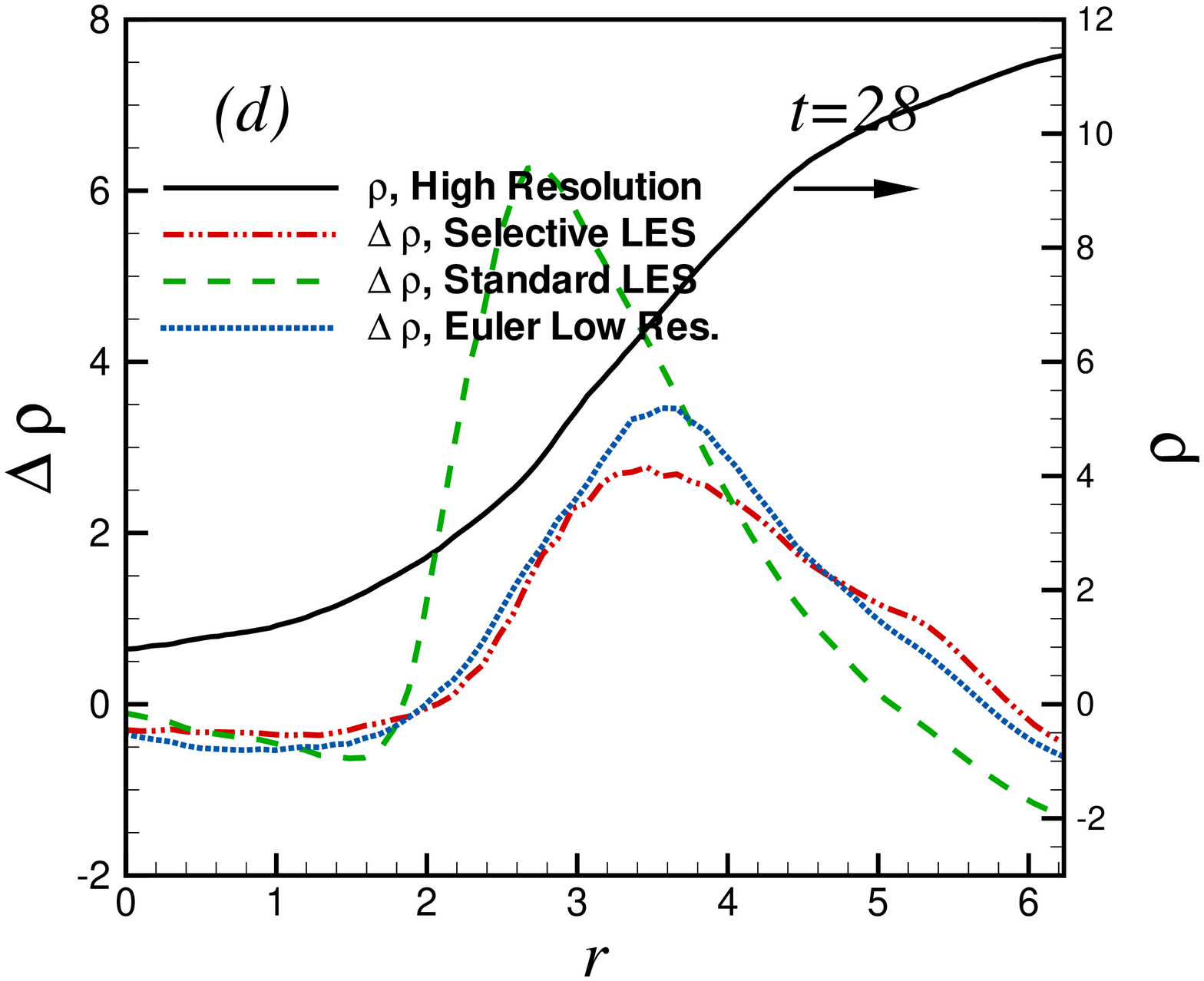}\\
\includegraphics[width=0.38\textwidth]{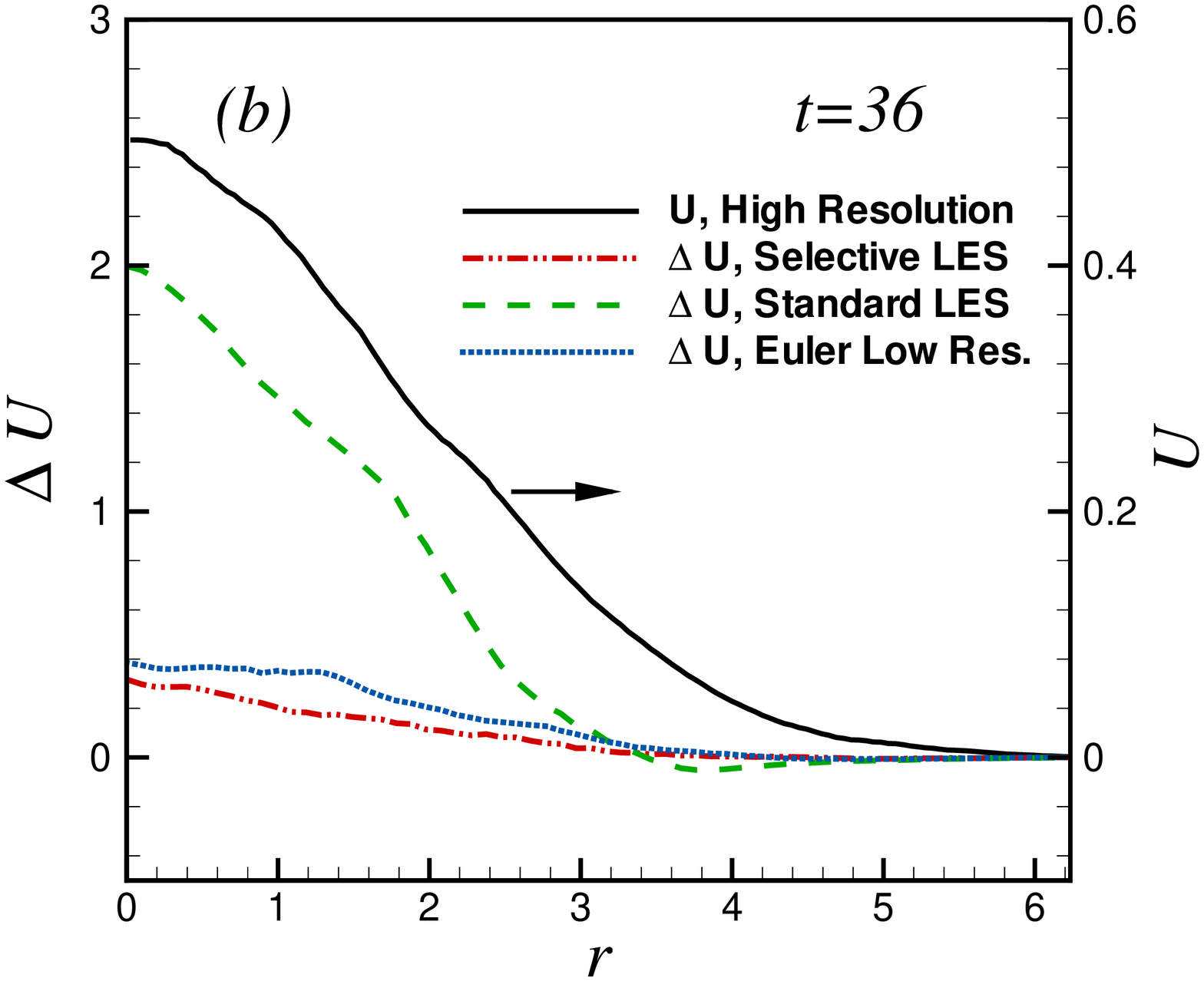}&
\includegraphics[width=0.38\textwidth]{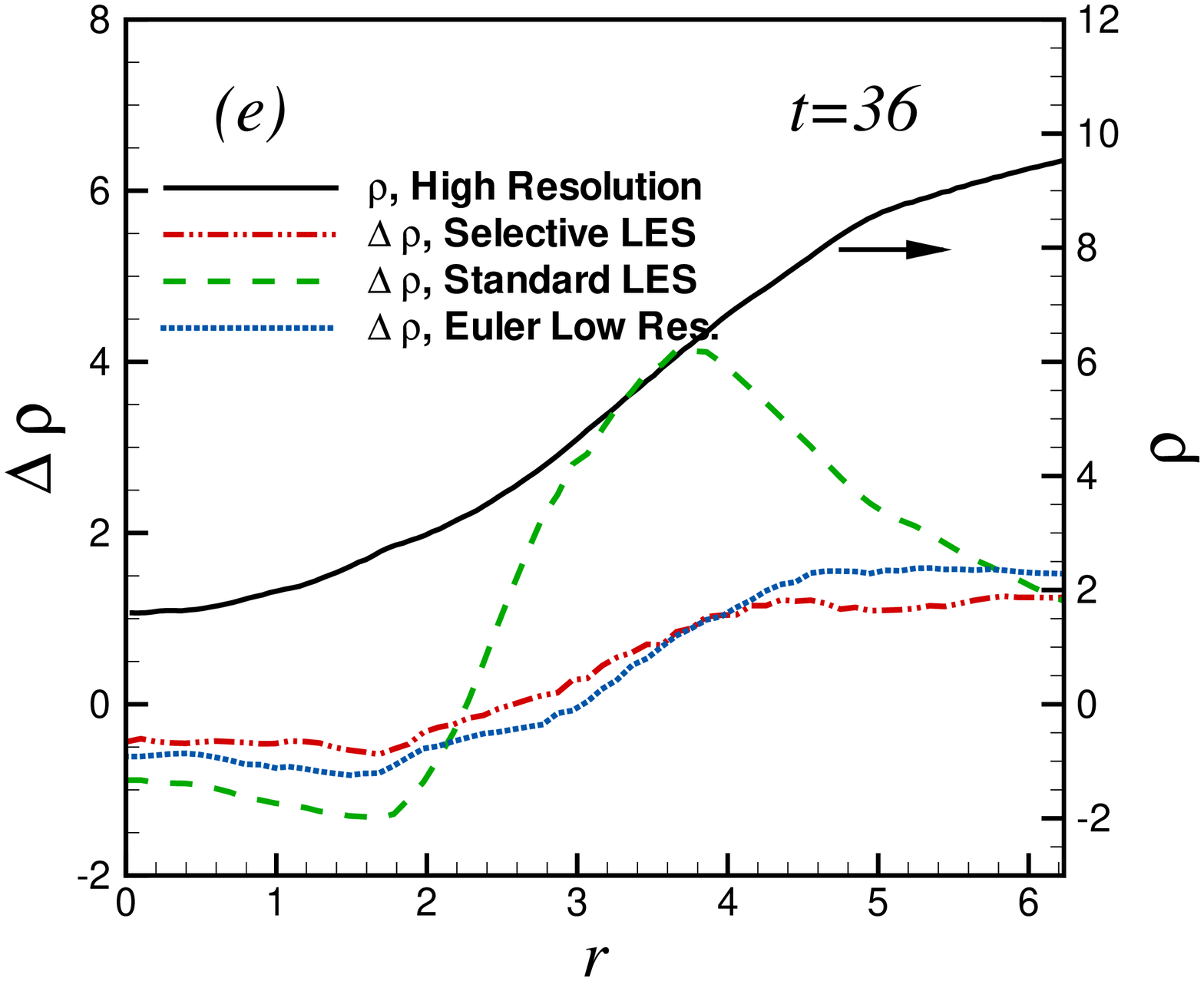}\\
\includegraphics[width=0.38\textwidth]{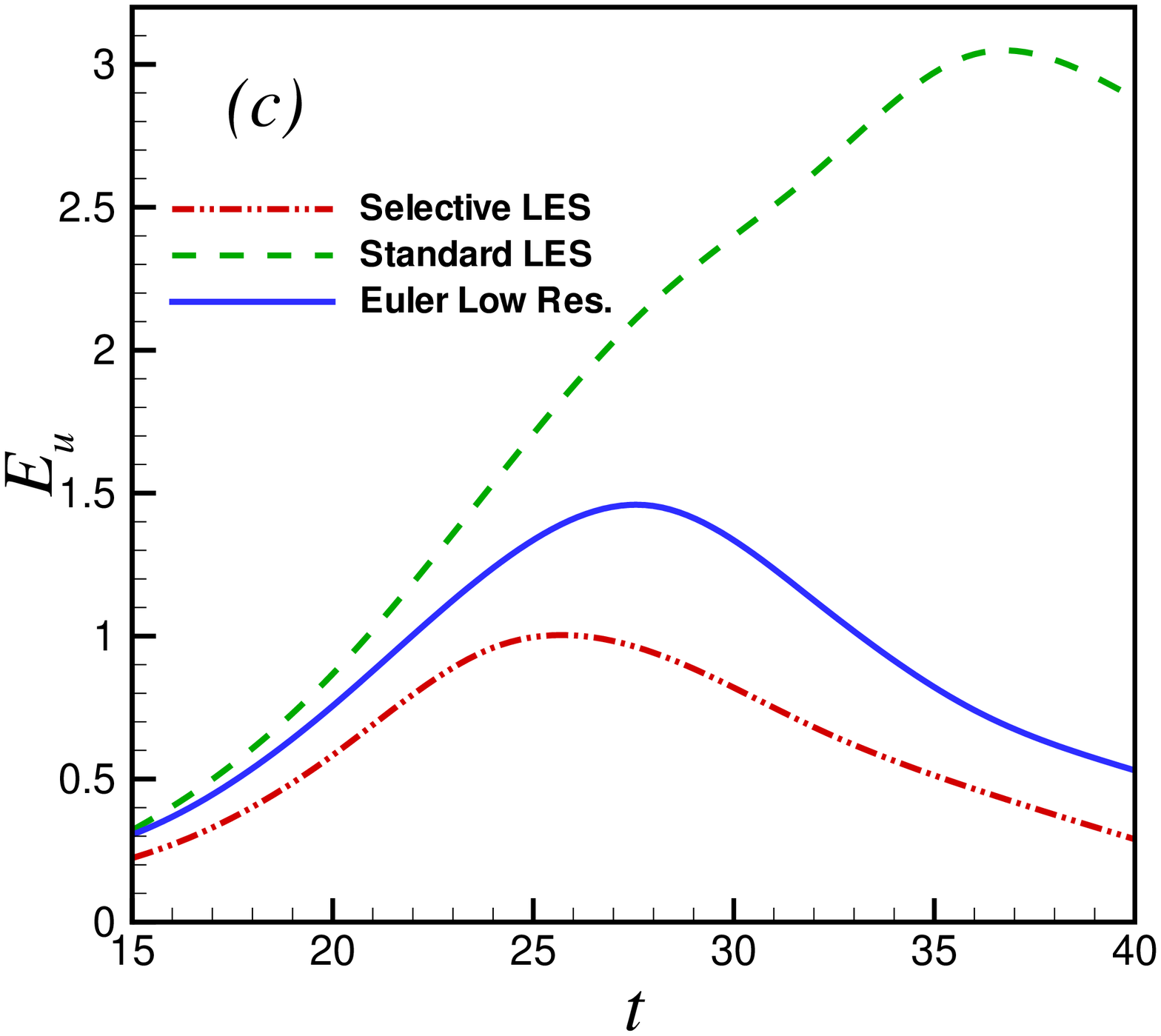}&
\includegraphics[width=0.38\textwidth]{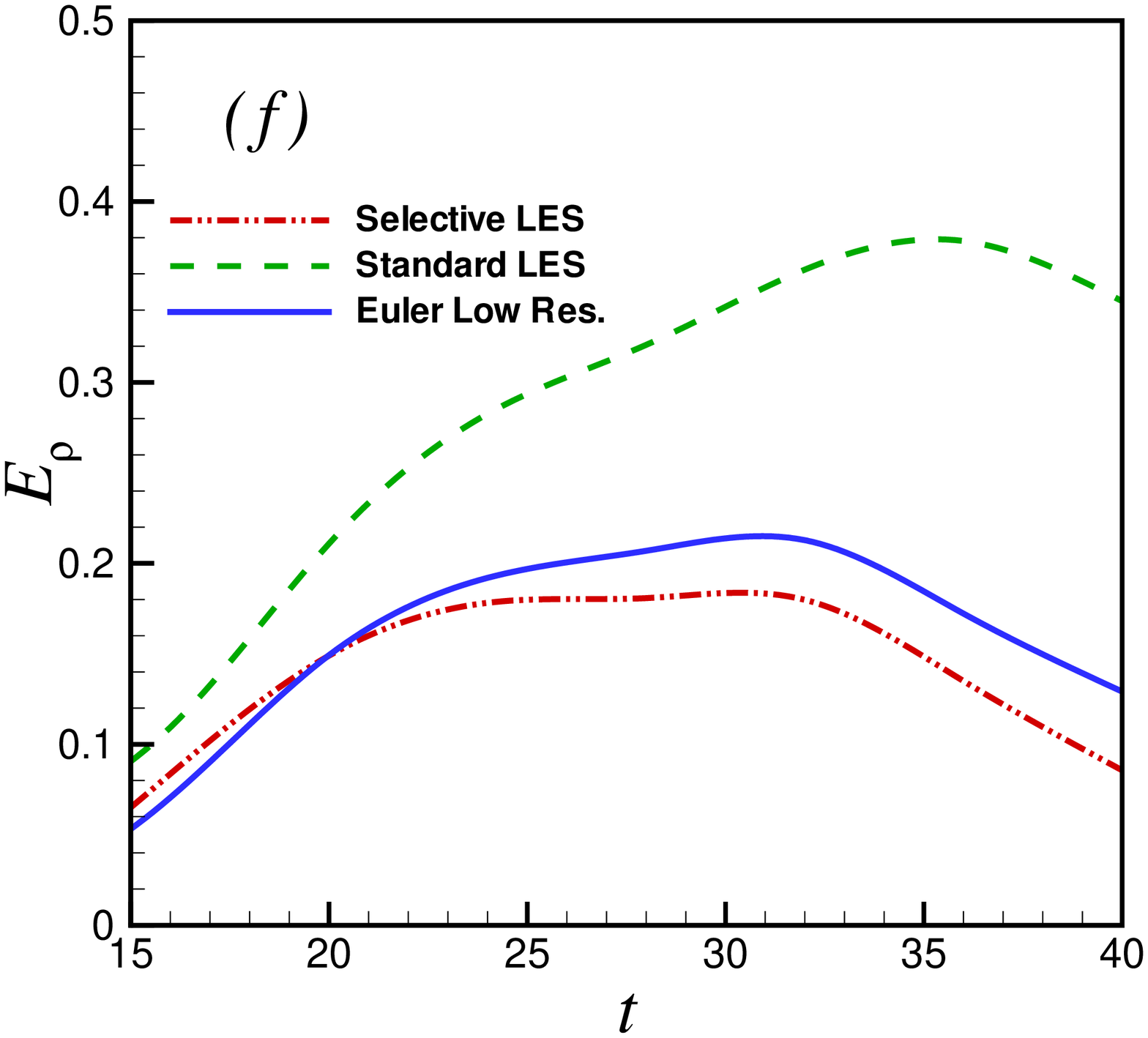}\\
\end{tabular}
\caption{Comparison between  the low and high resolution simulations. Radial distribution of the mean velocity difference, $\Delta U = U - U_{HR}$ (Panels a and b) and mean density difference, $\Delta \rho = \rho - \rho_{HR}$ (Panels d and e) at $t=28$ and $t=36$, where $U_{HR}$ is the high resolution mean velocity profile represented by the black lines in panels a and b and  $\rho_{HR}$ is the high resolution mean density profile represented by the black lines in panels d and e.
Panel c: normalized velocity difference $ E_u = \int_0^{2\pi R} \mid\Delta U\mid {\rm d}r / \int_0^{2\pi R} \ U_{HR}\, {\rm d}r$, Panel f: normalized density difference $ E_\rho = \int_0^{2\pi R} \mid\Delta \rho\mid {\rm d}r / \int_0^{2\pi R} \ \rho_{HR}\, {\rm d}r$, where $a$ is the initial jet radius.}
\label{fig.u-rho-media}
\end{figure*}

\begin{figure}
\centering
\includegraphics[width=0.89\columnwidth]{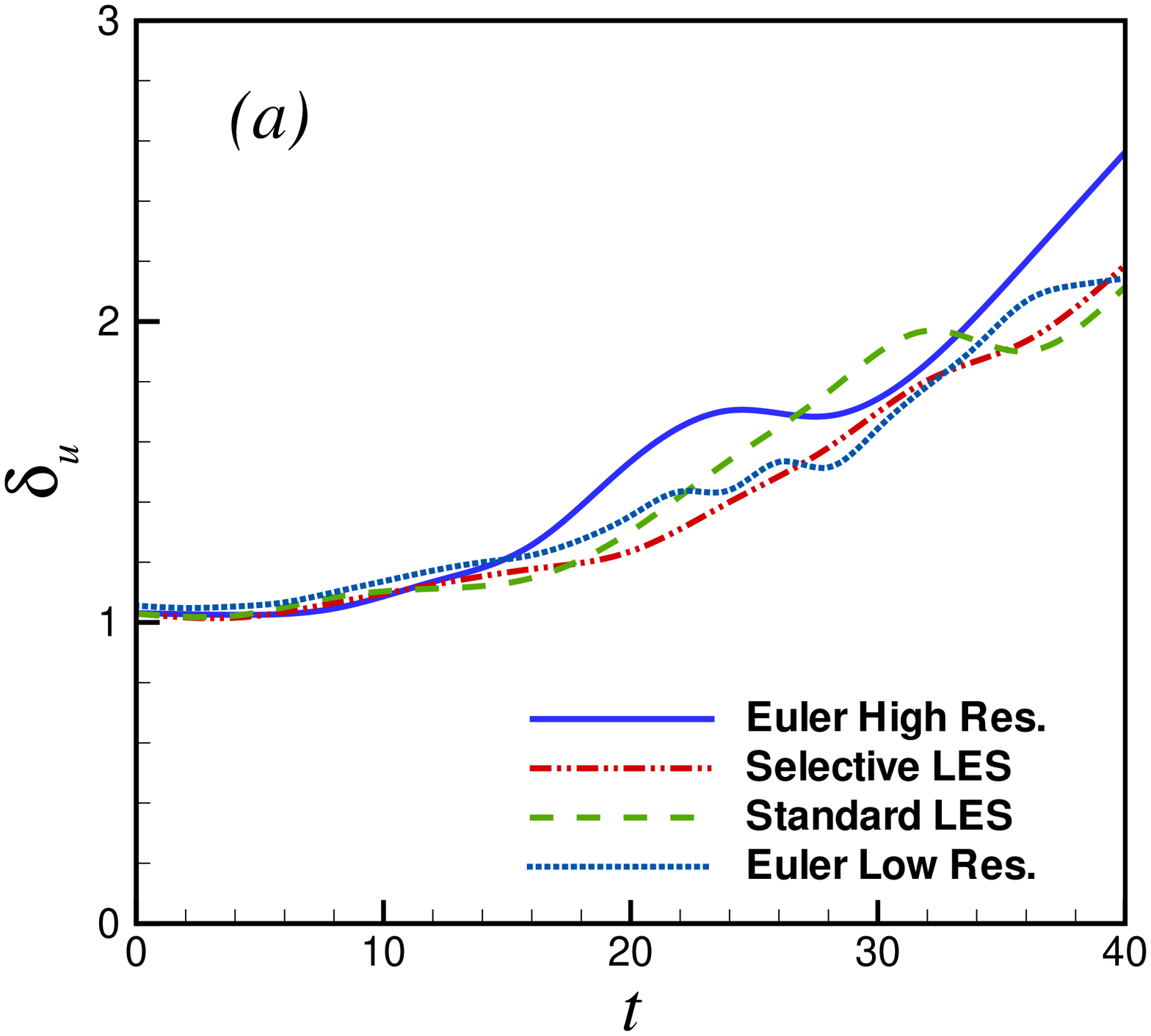}\\
\includegraphics[width=0.89\columnwidth]{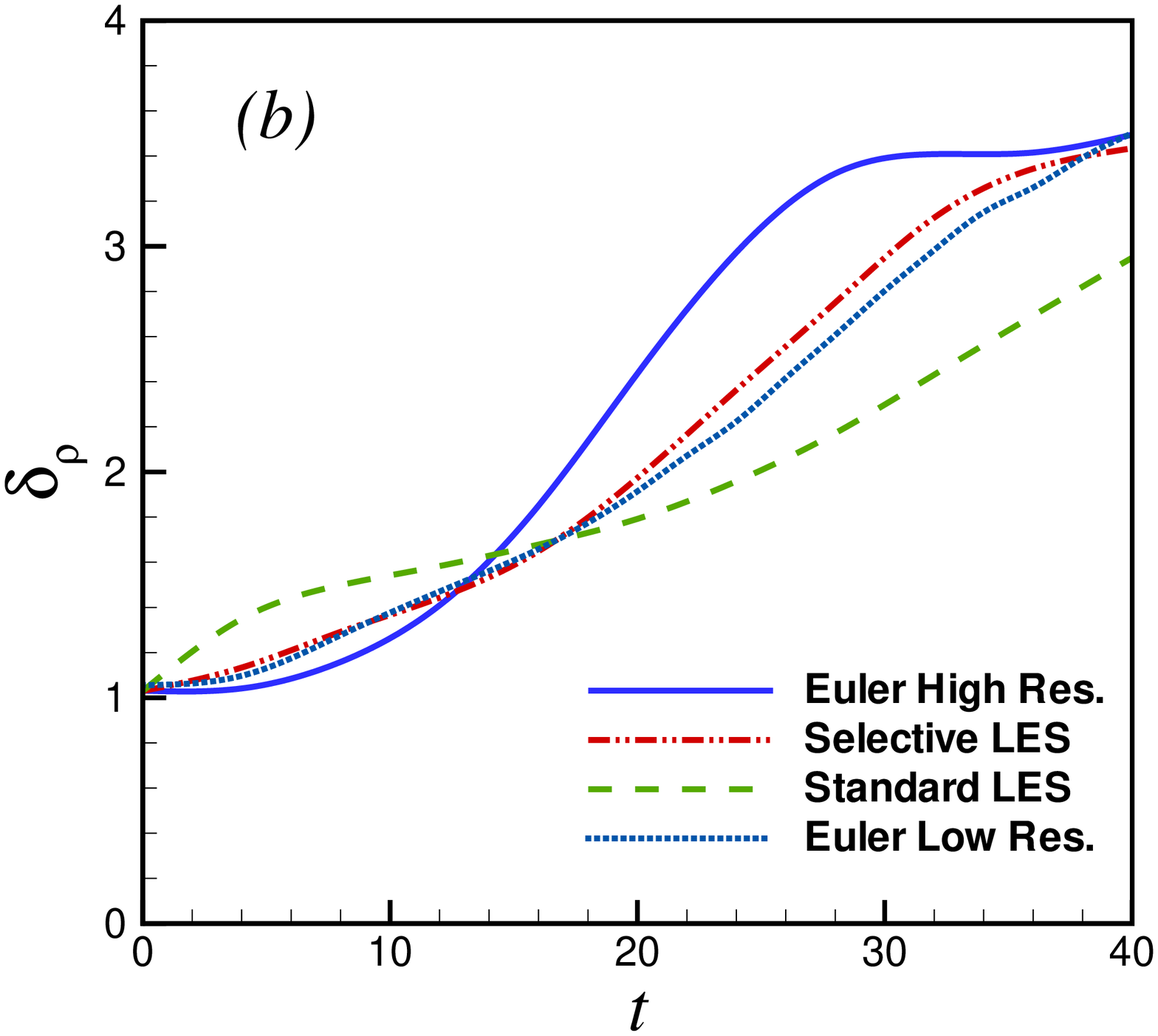}
\caption{Temporal evolution of the jet width: (a) velocity thickness $\delta_u$, defined as the distance between the jet axis and the position where the normalized mean velocity $U/U_0$ is equal to 0.5; (b) density thickness $\delta_\rho$, defined as the distance between the jet axis and the position where the mean density is equal to the average between the jet axis density and the external ambient density.}
\label{fig.delta-u-rho}-
\end{figure}

\begin{figure}
\centering
\includegraphics[width=\columnwidth]{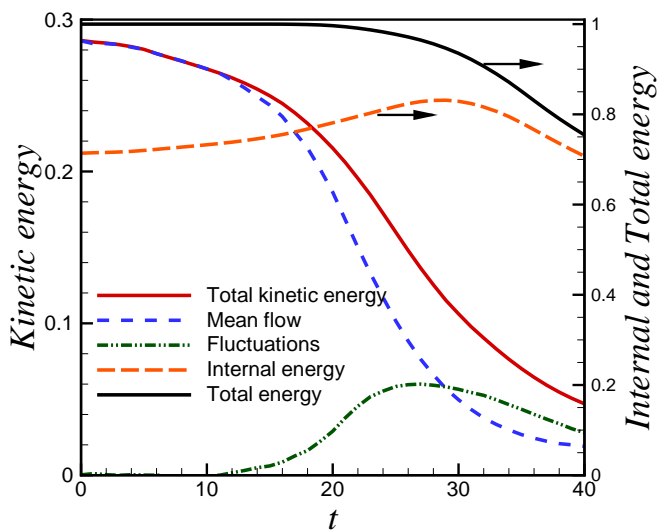}
\caption{Evolution of the mean kinetic energy $\overline{\rho u_k u_k}/2$ and internal energy $\overline{\rho e}$ in the computational domain, high resolution pseudo-DNS simulation. The kinetic energy has been decomposed in the sum of the energy of the mean flow $\overline{\rho}\tilde{u}_k\tilde{u}_k$ and of the energy of the fluctuations $\overline{\rho}\widetilde{u_k u_k}$. The tilde denotes density weighted Favre average: $\tilde u_k=\overline{\rho u_k}/\overline{\rho}$. All values have been normalized by the initial energy $\overline{\rho u_k u_k}/2 + \overline{\rho e}$. The evolution of the kinetic energy of the fluctuations determines the extension of the underresolved regions where subgrid terms must be introduced in the governing equations, see the movie in the supplementary material visible online.}
\label{fig.energia}
\end{figure}

A more quantitative assessment of the impact of the different modelling procedures on the overall flow features can be made by considering the mean quantities, in particular the velocity and density longitudinal distribution and the jet thickness.
All the lower resolution simulations tend to overestimate the velocity and the kinetic energy and to underestimate the density in the jet core as shown in figure \ref{fig.u-rho-media}, which shows the difference between the mean profiles obtained by the lower resolution simulations (selective LES, standard LES, Euler) and the higher resolution pseudo-DNS.
The overall better lower resolution simulation prediction is obtained by the selective LES, the worst by the standard LES. The Euler simulation performs worse than the selective LES, even if the difference between the two simulations tends to become smaller in the final stage of the simulation.
To evaluate the spreading rate of the jet, we consider the geometrical thickness $\delta_u$ of the velocity profiles, here defined as the distance from the jet axis where $\overline{u}/U_0 =0.5$, and the geometrical thickness $\delta_\rho$ of the density profiles, defined as the distance from the jet axis where $\overline{\rho} = (\overline{\rho}(0)+\overline{\rho}(\infty))/2$ (see figure \ref{fig.delta-u-rho}). While the geometrical velocity thickness $\delta_u$ does not seem to show a high sensitivity to the flow modelling, the density thickness clearly indicates the delay in the growth of the standard LES, which produces a reduced entrainment.

The temporal growth rate of the jet thickness can be transformed in an equivalent spatial growth rate by means of the Taylor transformation $x=U_0t$, that is,
$$
\frac{{\rm d}\delta}{{\rm d}x} = \frac{1}{U_0} \frac{{\rm d}\delta}{{\rm d} t}.
$$
In the first part of the simulation, up to $t=15$, the thickness grows very slowly. In this stage there the initial perturbations are still growing \cite{2000} and the mixing between the jet and external material is not yet begun.
The high resolution simulation has, in the second part of the simulation, an equivalent spatial growth rate of the velocity thickness $\delta_u$ equal to 0.029. The low resolution simulations begin the with a small delay and temporal growth rates are similar, the equivalent spatial growth rates are smaller: 0.025 for the selective LES, 0.023 for the Euler simulation and 0.013 for the standard LES. The high resolution simulation and the standard LES growth rates are in line with what can be expected in such a flow \cite{tim07,ps02}.
The growth rate of the density thickness $\delta_\rho$ is about 0.058 for the high resolution simulation, 0.052 for the selective LES, 0.022 for the standard LES and 0.049 for the Euler low resolution simulation. Part of the disagreement between the growth rates of the four simulations is due to the different longitudinal velocity $U_0$ predicted by the simulations.
The large delay in the growth of turbulence structures induced by the standard non selective LES, which is visible in the velocity and density spectra (see figures \ref{fig.spettri} and \ref{fig.spettri-rho}), is clearly evident also in the jet thickness, which presents a lower growth rate. This delay can be attributed to the overestimation of unresolved subgrid scales transport made by the standard non selective LES model, which leads to an initial damping of the resolved large scale structures which are mainly responsible for the jet entrainment.

As the turbulent energy associated with the fluctuation starts to decay, one can expect that the difference between the selective LES and the low resolution Euler simulation tends to become smaller. This can be observed  from the mean velocity and density profiles in figure \ref{fig.u-rho-media} and from the longitudinal spectra (figures \ref{fig.spettri} and \ref{fig.spettri-rho}). The depletion of the energy associated to the smaller scale motions in fact make the flow less turbulent.
As shown in figure \ref{fig.energia}, the kinetic energy of the velocity fluctuations reaches a maximum around $t=26$, becomes larger then the mean flow kinetic energy after $t=28$ but then decreases by about 40\%\ before the end of the simulation.
During this depletion of the jet energy, the volume of the regions where sub-grid scale terms are used is likewise reduced.
Therefore, it should be expected that the difference between the selective LES and the low resolution Euler simulation tends to be less relevant.


\section{Concluding remarks}
In this work we  show that the {\it selective} Large Eddy Simulation, which is based on the use of a scalar probe function $f$ -- a function of the magnitude of the local stretching-tilting term of the vorticity equation  -- can be conveniently applied to the simulation of time evolving compressible jets. In the present simulation, the probe function $f$ has been coupled with the standard Smagorinsky sub-grid model. However, it should be noted that the probe function $f$ can be used together with any model because $f$ simply acts as an independent switch for the introduction of a sub-grid model.
The main results is that even a simple model can give acceptable results when selectively used together with a sub-grid scale localization procedure. In fact, the comparison among the four kinds of simulations (selective LES, standard LES, low and high resolution pseudo Euler direct numerical simulations) here carried out shows that this method can improve the dynamical properties of the simulated field. In particular, the selective LES hugely improves the spectral distribution of energy and density over the resolved scales, the enstrophy radial distribution and the mean velocity (up to the 200\%) and density profiles (up to the 100\%) with respect to the standard LES. Furthermore, this method avoids the artificial over-damping of the unstable modes at the jet border which in the standard large eddy simulation inhibits the jet lateral growth.
In comparison with an Euler simulation which uses the same resolution, the selective LES clearly improves the flow prediction when the field is reach in small scales (up to the 50\% on the momentum and 4\% on the density fields). If, as in the example here shown, the kinetic energy in the small scales is not steady in the mean and decays,    the improvement due to the use of the selective LES in the long term reduces. Thus, in flow simulations where the small scales are transient in time these two methods asymptotically offer same results.

In synthesis, the selective LES explicitly introduces the sub-grid flows of momentum and energy in the governing equations in the regions of the flow where turbulence is physically present. In this way, one does not rely on the numerical diffusion to mimic the overall behaviour of all unresolved scales. This is a positive feature, since the numerical diffusion depends on the algorithm used and on the grid spacing and cannot  be conveniently controlled.

The computing time of the selective LES is about one third larger than that of the low resolution Euler simulation and seven times smaller than the one of the higher resolution Euler simulation. Therefore, a selective LES could be more convenient than a better resolved Euler simulation.
Because of this properties, given the modest computational burden brought to the simulation, the application of the selective procedure to the simulation of complex flows -- in particular highly compressible free flows as, for instance,  astrophysical jets -- seems promising.

\newpage


\begin{thebibliography}{99}
%

\bibitem{tim07}
D.Tordella, M.Iovieno, S.Massaglia,
``Small scale localization in turbulent flows. A priori tests applied to a possible Large Eddy Simulation of compressible turbulent flows'',
{\it Comp.\ Phys.\ Comm.} {\bf 176}(8), 539--549 (2007).


\bibitem{lilly}
D.K.Lilly
``The representation of small-scale turbulence in numerical simulation experiments'',
  {\itshape Proc.\ IBM Scientific Computing Symp.\ on Environmental Sciences}, Yorktown Heights, New York, ed.\
  H.H.\ Goldstine, IBM form no.\ 320--1951 (1967).

\bibitem{cc97}
V.M.Canuto, Y.Cheng,
``Determination of the Smagorinsky--Lilly constant $C_s$'',
{\itshape Phys.\ Fluids} {\bf 9}(5), 1368--1378 (1997).



\bibitem{toschi}
L.Biferale, G.Boffetta, A.Celani, A.Lanotte, F.Toschi,
``Particle trapping in three-dimensional fully developed turbulence'',
{\it Phys.\ Fluids.} {\bf 17}(2), 021701 (2005).


\bibitem{morkovin}
M.V.Morkovin,
``Effects of compressibility on turbulent flows'',
in {\it M\`ecanique de la turbulence}, edited by A.\ Favre, 367, (1961).

\bibitem{br}
G.L.Brown, A.Roshko,
``On density effects and large structure in turbulent mixing layers'',
{\it J.\ fluid Mech.} {\bf 64}, 775--816, (1974).

\bibitem{ps02}
C.Pantano, S.Sarkar,
``A study in compressibility effects in the high-speed shear flows using direct simulations'',
{\it J.Fluid Mech.},{\bf 451}, 329--371 (2002).


\bibitem{germano}
M.Germano, U.Piomelli, P.Moin, W.H.Cabot, ``A dynamic subgrid-scale eddy viscosity model'', {\itshape Phys.\ Fluids A} {\bf 3}, 1760, (1991).

\bibitem{vreman}
A.W.Vreman,
``An eddy-viscosity subgrid-scale model for turbulence shear flow: algebraic theory and applications'',
{\it Phys.\ Fluids} {\bf 16}, 3670--3681, (2004).

\bibitem{sak01}
S.Stolz, N.A.Adams, L.Kleiser,
``The approximate deconvolution model for the large-eddy simulation of compressible flows and its application to shock-turbulent boundary layer interaction'',
{\it Phys.Fluids} {\bf 10}, 2985--3001, (2001).





\bibitem{be99}
F.Bacciotti, J.Eisl\"offel,
``Ionization and density along the beams of Herbig-Haro jets'',
{\itshape Astrononomy and Astrophys.} {\bfseries 342}, 717--735 (1999).


\bibitem{rb01}
B.Reipurth, J.Bally,
``Herbig-Haro flows: Probes of early stellar evolution'',
{\itshape Annual Review of Astronomy and Astrophysics} {\bfseries 39}, 403--455 (2001).

\bibitem{hardee00}
J.M.Stone, P.E.Hardee and J.Xu,
``The stability of radiatively cooled jets in three dimensions'',
{\itshape Astrophysical Journal} {\bf 543}(1), 161--167 (1997). 


\bibitem{rossi97}
P.Rossi, G.Bodo, S.Massaglia, A.Ferrari,
``Evolution of Kelvin-Helmholtz instabilities in radiative jets .2. Shock structure and entrainment properties'',
{\itshape Astronomy and Astrophysics} {\bf 321}(2), 672--684 (1997). 



\bibitem{pluto}
A.Mignone, G.Bodo , S.Massaglia, T.Matsakos, O.Tesileanu, C.Zanni and A.Ferrari,
``PLUTO: a numerical code for computational astrophysics'',
{\it Astr.\ J. Supplement Series} {\bf 170}(1), 228--242 (2007),
and http://plutocode.to.astro.it.

\bibitem{cw84}
P.Colella, P.R.Woodward,
``The piecewise parabolic method (PPM) for gas-dynamical simulations,
{\it J.\ comp.\ Phys.\ }{\bf 54}(1), 174--201, (1984).

\bibitem{d99}
F.Ducros, V.Ferrand, F.Nicoud, C.Weber, D.Darraq, C.Gacherieu, T.Poinsot,
``Large-eddy simulation of the high-turbulence interaction'',
{\it J.\ comp.\ Phys.} {\bf150}, 199--238, (1999).



\bibitem{njp2011}
D.Tordella, M.Belan, S.Massaglia, S.De Ponte, A.Mignone, E.Bodenschatz, A.Ferrari, ``Astrophysical jets: insights into long-term hydrodynamics'', {\itshape New Journal of Physics} {\bf 13}, 043011 (2011).



\bibitem{it03}
M.Iovieno, D.Tordella,
``Variable scale filtered Navier-Stokes equations: A new procedure to deal with the associated commutation error'',
{\itshape Phys.\ Fluids} {\bf 15}(7), 1926--1936 (2003). 

\bibitem{2000}
M.\ Micono, G.\ Bodo, S.\ Massaglia, P.\ Rossi, A.\ Ferrari, R.\ Rosner,
``Kelvin-Helmholtz Instabilities in Three Dimensional Radiative Jets'',
{\it Astronomy and Astrophys.}, {\bf 360}, 795--808 (2000).



\bibitem{bm98}
G.Bodo, P.Rossi, S.Massaglia,
``Three-dimensional simulations of jets'',
{\it Astronomy and Astrophys.} {\bf 333}, 1117--1129 (1998).








\bibitem{hardee95}
P.E.Hardee, D.A.Clarke and D.A.Howell,
``The stability and collimation of three dimensional jets'',
{\itshape Astrophysical Journal} {\bf 441}(2), 644--664 (1995). 



\bibitem{dr98}
T.P.Downes, T.P.Ray,
``Numerical simulations of the Kelvin-Helmholtz instability in radiatively cooled jets'',
{\itshape Astronomy and Astrophysics} {\bf 331}(3), 1130--1142 (1998). 



\bibitem{hardee88}
P.E.Hardee, M.L.Norman
``Spatial stability of the slab jet. I - Linearized stability analysis, II - Numerical simulations.'',
{\itshape The Astrophysical Journal} {\bf 334}, 70--94 (1988). 



\bibitem{bm94}
G.Bodo, S.Massaglia, A.Malagoli, A.Ferrari, E.Trussoni,
``Kelvin-Helmholtz instability of hydrodynamic supersonic jets'',
{\itshape Astronomy and Astrophysics} {\bf 283}(1), 655--676 (1994). 



\bibitem{bodo95}
G.Bodo, S.Massaglia, P.Rossi, R.Rosner, A.Malagoli, A.Ferrari,
``The long-term evolution and mixing properties of high Mach number hydrodynamic jets'',
{\itshape Astronomy and Astrophysics} {\bf 303}(1), 281--298 (1995). 





\end{thebibliography}
\end{document}